\documentclass[sigconf]{acmart}
\usepackage{multirow}
\usepackage{makecell}
\usepackage{xcolor}
\usepackage{array}
\usepackage{booktabs}
\usepackage{tabularx}
\usepackage{threeparttable}
\usepackage{graphicx}
\usepackage{tcolorbox}
\usepackage{CJKutf8}
\usepackage[table]{xcolor} 
\usepackage[shortcuts]{extdash}
\definecolor{tableblue}{RGB}{235, 245, 255}

\renewcommand\footnotetextcopyrightpermission[1]{} 
\AtBeginDocument{%
  }

\setcopyright{none}
\copyrightyear{2026}
\acmYear{2026}
\acmDOI{}
\acmConference[Preprint]{}
\acmBooktitle{}
\acmISBN{}

\makeatletter
\renewcommand\@author{}
\makeatother

\begin{document}

\title{MINT-Bench: A Comprehensive Multilingual Benchmark for Instruction-Following Text-to-Speech}

\author{
  \Large{\textbf{
    Huakang Chen$^{1,4}$, 
    Jingbin Hu$^{1}$, 
    Liumeng Xue$^{2}$,
    Qirui Zhan$^{1}$,
    Wenhao Li$^{1}$,
    Guobin Ma$^{1}$,
    Hanke Xie$^{1}$,
    Dake Guo$^{1}$,
    Linhan Ma$^{1}$,
    Yuepeng Jiang$^{1}$,
    Bengu Wu$^{3}$,
    Pengyuan Xie$^{4}$,
    Chuan Xie$^{4}$,
    Qiang Zhang$^{4}$,
    Lei Xie$^{1,*}$
  }}\\
  \vspace{8pt}
  $^{1}$Audio, Speech and Language Processing Lab (ASLP@NPU), \\ Northwestern Polytechnical University, Xi'an, China\\
  $^{2}$School of Intelligence Science and Technology, \\ Nanjing University, Suzhou, China\\
  $^{3}$Yutu Zhineng, Beijing, China\\
  $^{4}$Lingguang Zhaxian Technology, Shanghai, China\\
  \vspace{5pt}
  *Corresponding author
}

\renewcommand{\shortauthors}{Chen et al.}

\begin{abstract}
Instruction-following text-to-speech (TTS) has emerged as an important capability for controllable and expressive speech generation, yet its evaluation remains underdeveloped due to limited benchmark coverage, weak diagnostic granularity, and insufficient multilingual support. We present \textbf{MINT-Bench}, a comprehensive multilingual benchmark for instruction-following TTS. MINT-Bench is built upon a hierarchical multi-axis taxonomy, a scalable multi-stage data construction pipeline, and a hierarchical hybrid evaluation protocol that jointly assesses content consistency, instruction following, and perceptual quality. Experiments across ten languages show that current systems remain far from solved: frontier commercial systems lead overall, while leading open-source models become highly competitive and can even outperform commercial counterparts in localized settings such as Chinese. The benchmark further reveals that harder compositional and paralinguistic controls remain major bottlenecks for current systems. We release MINT-Bench together with the data construction and evaluation toolkit to support future research on controllable, multilingual, and diagnostically grounded TTS evaluation. The leaderboard and demo are available at \url{https://aslp-lab.github.io/MINT-Bench-Demo/}.
\end{abstract}


\maketitle

\section{Introduction}
\label{sec:intro}

Text-to-speech (TTS) has advanced rapidly in recent years ~\cite{valle, maskgct, f5tts, llasa, sparktts, vibevoice, voxcpm, vevo2, xie2025soulxpodcastrealisticlongformpodcasts, kalle, ditar, liao2026fishaudios2technical}, evolving from a technology primarily judged by intelligibility and speaker clarity into one that is increasingly natural, expressive, and controllable. In modern applications, synthesized speech is expected not only to deliver the correct verbal content, but also to convey it in the requested way---with appropriate timbre, speaking style, emotion, persona, or other vocal characteristics. This trend has driven TTS beyond plain text-conditioned synthesis toward \emph{instruction-following} generation, where users specify desired vocal behaviors through natural language instructions rather than fixed symbolic controls.

This shift substantially expands the practical scope of speech generation. Compared with traditional tag-based control, natural language provides a more flexible and user-friendly interface for specifying vocal attributes at different levels of abstraction, ranging from atomic acoustic properties to compositional, scenario-based, and dynamically evolving behaviors. As a result, instruction-following TTS is becoming a core capability for applications such as storytelling, conversational agents, role-play, and personalized audio content ~\cite{unistyle, histyle, cosyvoice2, flespeech, indextts2, indextts2_5, qwen3tts}. Recent open-source and commercial systems have increasingly moved in this direction, making reliable evaluation of instruction-following ability an urgent problem for both research and deployment.

Evaluating instruction-following TTS, however, is significantly more challenging than evaluating conventional TTS. A successful TTS system must simultaneously preserve the intended linguistic content, realize the requested control target, and maintain convincing perceptual quality ~\cite{fireredtts, cosyvoice, indextts, qwen3tts} . These aspects are related but not interchangeable: content consistency is the prerequisite, instruction following is the core capability, and perceptual quality determines whether the generated speech is practically usable. Moreover, speech is inherently one-to-many: the same instruction may admit multiple valid realizations, especially when the target involves open-ended style, persona, or extra-vocal expression. As a result, evaluation cannot be reduced to either transcript fidelity alone or a single aggregate quality score.

Existing evaluation resources remain insufficient for this setting. Traditional TTS metrics ~\cite{seed-tts-eval} mainly focus on intelligibility, acoustic similarity, or overall naturalness, and are not well-suited to open-ended instruction following. Existing instruction-following benchmarks ~\cite{instruct-tts-eval, s2s-arena} have made encouraging progress, but several gaps remain. First, benchmark coverage is often limited or weakly organized, especially for timbre-related control, richer compositional cases, and paralinguistic or extra-vocal behaviors. Second, existing protocols often provide limited diagnostic granularity, making it difficult to disentangle whether a system fails because of content errors, weak execution of the requested instruction, or degraded perceptual quality. Third, many benchmarks rely on relatively coarse judgments of instruction adherence and offer limited support for systematic multilingual expansion. Taken together, these limitations suggest the need for a benchmark that is simultaneously comprehensive in coverage, diagnostic in evaluation, and scalable in construction.

To address these challenges, we present \textbf{MINT-Bench}, a comprehensive \textbf{M}ultilingual benchmark for \textbf{IN}struction-following \textbf{T}TS. Rather than treating evaluation as a flat collection of prompts, MINT-Bench formulates it as a structured and extensible benchmark construction problem. 

At its core is a \emph{Hierarchical Multi-axis Taxonomy} grounded in ten atomic timbre and style attributes that define the core benchmark inventory. Based on this inventory, MINT-Bench covers structured variants of timbre and style control, including composition, dynamic or layered realization, persona- or scenario-based prompting, and \emph{Special} cases involving abnormal vocalization and nonverbal events. 
Building on this taxonomy, we construct a scalable multi-stage data construction pipeline that generates semantically clear instruction-text pairs and can be naturally extended to new control settings and languages. MINT-Bench further introduces a \emph{hierarchical hybrid evaluation protocol} that progressively evaluates \emph{content consistency}, \emph{instruction following}, and \emph{perceptual quality} by combining objective tools with Large Audio Language Model (LALM)-based judgments ~\cite{emergenttts}. In this way, MINT-Bench provides not only overall ranking signals, but also fine-grained diagnosis of where systems succeed or fail. 

Experiments across ten languages show that instruction\-/following TTS remains far from solved: although strong commercial systems currently define the overall performance frontier, leading open-source systems have become highly competitive and can even surpass commercial counterparts in localized settings such as Chinese. The benchmark further reveals that strong content fidelity does not necessarily imply strong controllability, and that harder compositional and \emph{Special} extra-vocal cases remain major bottlenecks for current systems, highlighting the importance of going beyond flat prompt sets and aggregate scores toward a more structured and diagnostically informative evaluation framework.

The key contributions of our work are summarized as follows:

\begin{itemize}
    \item We present \textbf{MINT-Bench}, a comprehensive multilingual benchmark for instruction-following TTS designed to support systematic, fine-grained, and scalable evaluation of controllability and generation quality. We will open-source MINT-Bench, together with its data construction and evaluation toolkit, to facilitate future research on instruction-following TTS.

    \item We introduce a hierarchical multi-axis taxonomy together with a scalable multi-stage data construction pipeline, enabling semantically clear, structurally controlled, and extensible benchmark construction across diverse control settings and languages.

    \item We design a hierarchical hybrid evaluation protocol that progressively assesses content consistency, instruction following, and perceptual quality by combining objective tools with LALM-based judgments.
\end{itemize}

\begin{figure*}[h]
  \centering
  \includegraphics[width=0.9\linewidth]{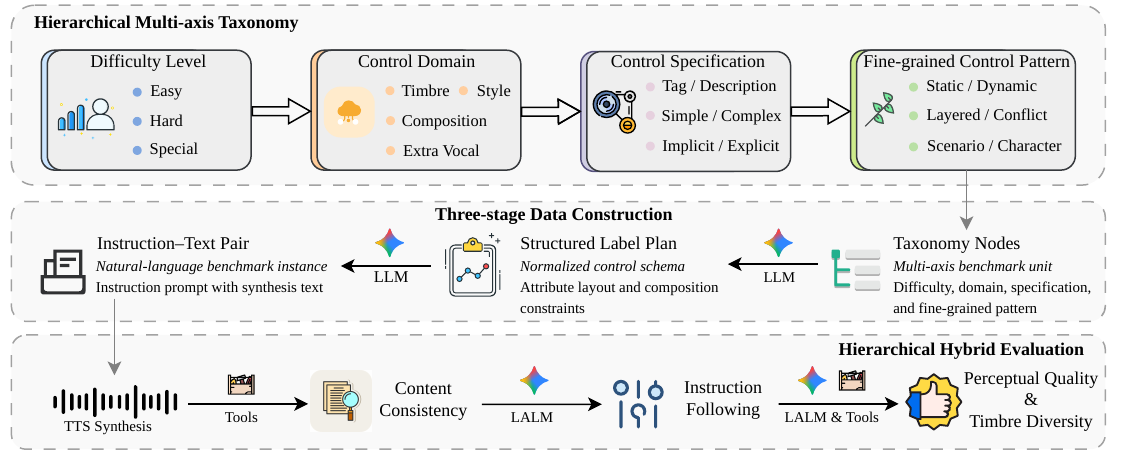}
  \caption{Overview of MINT-Bench, consisting of a hierarchical multi-axis taxonomy, a three-stage data construction pipeline, and a hierarchical hybrid evaluation protocol.}
  \label{fig:overview}
\end{figure*}

\section{Related Work}
\label{sec:related_work}

\subsection{Controllable and Instruction-Following TTS}

Controllable TTS has evolved from structured attribute manipulation toward open-ended natural-language-guided generation. Early controllable systems typically relied on predefined tags, special tokens, or other structured control signals to regulate attributes such as speaker identity, emotion, and prosody. Representative systems such as ST-TTS ~\cite{st-tts} , CosyVoice ~\cite{cosyvoice} , FireRedTTS ~\cite{fireredtts} and SparkTTS ~\cite{sparktts} show that such interfaces can provide effective and parameterizable control within a closed and relatively well-defined attribute space. However, their flexibility is fundamentally bounded by the coverage and granularity of the underlying control inventory, which makes them less suitable for real-world user instructions that are often diverse, underspecified, and compositionally structured.

More recent systems increasingly adopt natural language as the control interface, allowing users to specify desired vocal characteristics through free-form textual descriptions rather than fixed symbolic tags. This shift has led to the emergence of instruction-following TTS, where the model is expected to interpret and realize heterogeneous requests involving timbre, speaking style, persona, emotion, and other expressive factors. Representative open-source systems include InstructTTS ~\cite{instruct-tts} , the PromptTTS family ~\cite{prompttts, prompttts++, prompttts2} , PromptStyle ~\cite{promptstyle} , Parler-TTS ~\cite{parler} , CosyVoice~2 ~\cite{cosyvoice2} , VoxInstruct ~\cite{voxinstruct} , EmoVoice ~\cite{emovoice} , Qwen3-TTS ~\cite{qwen3tts} , IndexTTS2 ~\cite{indextts2} , VoiceSculptor ~\cite{hu2026voicesculptorvoicedesigned} , MOSS-VoiceGenerator ~\cite{huang2026mossvoicegeneratorcreaterealisticvoices} , and Ming-omni-tts \footnote{\url{https://github.com/inclusionAI/Ming-omni-tts}} , while commercial systems such as Gemini ~\cite{gemini2.5}, ElevenLabs \footnote{\url{https://elevenlabs.io/app/api/eleven_ttv_v3}} , GPT-4o-mini-TTS \footnote{\url{https://developers.openai.com/api/docs/models/gpt-4o-mini-tts}} , MiniMax-Speech \footnote{\url{https://api.minimaxi.com/v1/voice_design}} , and Hume \footnote{\url{https://www.hume.ai/}} demonstrate that natural-language-driven speech control has already become relevant in deployed products.

This transition from closed-set control to open-ended instruction substantially changes the evaluation target. In contrast to tag-based systems that mainly test whether a predefined attribute can be activated, instruction-following TTS must handle richer forms of control, including compositional descriptions, dynamic or layered changes, persona- or scenario-based prompts, and extra-vocal behaviors. As a result, evaluation requires not only broader benchmark coverage, but also finer diagnostic granularity to reveal which kinds of control a system can or cannot execute reliably.

\subsection{Benchmarks and Evaluation for Instruction-Following TTS}

Existing resources for controllable or prompt-based speech generation provide useful foundations, but they only partially address the benchmark needs of instruction-following TTS. Early datasets such as TextrolSpeech ~\cite{TextrolSpeech} and ParaSpeechCaps ~\cite{ParaSpeechCaps}  mainly provide style-related annotations or promptable speech resources. These resources are valuable for training and initial probing, but they do not by themselves define a comprehensive evaluation framework for systematically measuring multidimensional controllability across heterogeneous instruction types.

More recent efforts have begun to move toward structured evaluation. InstructTTSEval ~\cite{instruct-tts-eval} introduces a more organized automatic evaluation setting for instruction-following TTS, while related efforts such as VStyle ~\cite{vstyle} and Speech-DRAME ~\cite{speech-drame} explore broader speech-native or human-aligned evaluation paradigms for expressive spoken generation. These works offer important insights, but their scope remains limited in different ways. InstructTTSEval is closer to general instruction-following evaluation, yet its diagnostic factorization over diverse control patterns is still limited. VStyle covers a relatively broad set of acoustic attributes and implicit empathy, but is primarily designed for Spoken Language Models rather than general-purpose TTS and relies heavily on a single LALM-based judge. Speech-DRAME emphasizes role-play-oriented, speech-native evaluation, but focuses on a more specialized application setting rather than general multi-axis control in TTS.

Overall, existing evaluation resources still leave three key gaps. First, benchmark coverage is often insufficiently structured, especially for timbre-related control, richer compositional cases, and extra-vocal behaviors. Second, existing protocols typically offer limited diagnostic granularity, making it difficult to disentangle content consistency, instruction following, and perceptual quality. Third, many existing resources remain limited in multilingual extensibility. MINT-Bench is designed to address these gaps through a hierarchical multi-axis taxonomy, a scalable multi-stage data construction pipeline, and a hierarchical hybrid evaluation protocol, thereby providing a more comprehensive, diagnostic, and multilingual benchmark for instruction-following TTS.

\section{MINT-Bench}
\label{sec:mint_bench}

\subsection{Overview}
\label{sec:overview}

Figure~\ref{fig:overview} presents the overall framework of MINT-Bench. Rather than treating instruction-following TTS evaluation as a flat collection of prompts, MINT-Bench formulates it as a structured benchmark construction and evaluation problem. The framework consists of three tightly coupled components: a \emph{Hierarchical Multi-axis Taxonomy} that organizes benchmark coverage, a controlled data construction pipeline that instantiates valid benchmark cases into natural-language items, and a hierarchical hybrid evaluation protocol that assesses synthesized speech from multiple complementary perspectives.

\begin{table}[t]
\centering
\small
\setlength{\tabcolsep}{10pt} 
\renewcommand{\arraystretch}{1.3}

\caption{Ten atomic acoustic attributes that form the core timbre and style inventory of MINT-Bench.
}
\label{tab:attributes}
\resizebox{\columnwidth}{!}{
\begin{tabular}{lll}
\toprule
\bfseries Category & \bfseries Attribute & \bfseries Description / Levels \\
\midrule

\multirow{4}{*}{\bfseries Timbre}  & Age         & 3 classes (Child, Adult, Elderly) \\
                                   & Gender      & 2 classes (Male, Female) \\
                                   & Pitch       & 5-level perceived pitch \\
                                   & Texture     & Human-annotated (e.g., Magnetic, Smoky, Nasal) \\
\midrule
\multirow{6}{*}{\bfseries Style}   & Speed       & 5-level speaking rate \\
                                   & Volume      & 5-level perceived loudness \\
                                   & Emotion     & 7 standard classes (e.g., Happy, Angry, Fearful) \\
                                   & Accent      & Human-annotated (e.g., Beijing Accent, British Accent) \\
                                   & Personality & Human-annotated (e.g., Humorous, Gentle, Aloof) \\
                                   & Tone        & Human-annotated (e.g., Firm, Awkward, Disappointed) \\
\bottomrule
\end{tabular}
}
\end{table}

\begin{table*}[t]
\centering
\footnotesize 
\setlength{\tabcolsep}{3pt}
\renewcommand{\arraystretch}{1.05} 

\caption{Representative examples of the MINT-Bench data construction pipeline, showing how taxonomy nodes are converted into structured label plans and then realized as final instruction-text pairs across \textit{Easy}, \textit{Hard}, and \textit{Special} cases.}
\label{tab:data_construction_examples}
\resizebox{\textwidth}{!}{ 
\begin{tabular}{p{0.15\textwidth} p{0.35\textwidth} p{0.25\textwidth} p{0.20\textwidth}}
\toprule
\textbf{Taxonomy Node} & \textbf{Structured Label Plan} & \textbf{Instruction} & \textbf{Text} \\
\midrule

\textbf{Easy / Timbre / Direct / Atomic / Age}
&
\textbf{Target:} age = child \newline
\textbf{Form:} direct \newline
\textbf{Text semantic:} simple request \newline
\textbf{Text length:} short \newline
\textbf{Key constraints:} natural phrasing; avoid child-themed lexical leakage
&
Speak as if you are a young child.
&
Can I have a glass of juice, please? \\

\midrule

\textbf{Hard / Composite / Complex / Dynamic}
&
\textbf{Targets:} emotion = thoughtful $\rightarrow$ angry; 
speed = slow $\rightarrow$ fast; 
volume = quiet $\rightarrow$ very\_loud \newline
\textbf{Form:} direct \newline
\textbf{Text semantic:} explanation / realization \newline
\textbf{Text length:} long \newline
\textbf{Key constraints:} staged progression; enough content for dynamic transitions
&
Begin speaking slowly, quietly, and thoughtfully. As you explain the situation, build your emotion into a fast, very loud, and angry delivery.
&
At first, I thought it was just a simple mistake. A clerical error. But then I saw the pattern, how it affected only certain people, and I realized this wasn't an accident. It was a deliberate choice, and it is absolutely unacceptable! \\

\midrule

\textbf{Special / Extra\_Vocal / Implicit\_NV}
&
\textbf{Target:} implicit\_nv = cough \newline
\textbf{Aux:} placement = interspersed; frequency = multiple times \newline
\textbf{Form:} direct \newline
\textbf{Text semantic:} explanatory sentence \newline
\textbf{Text length:} medium \newline
\textbf{Key constraints:} no explicit tags; text must not be about illness
&
As you're explaining this, interrupt yourself with a couple of small coughs, like you're trying to clear your throat.
&
It's important to understand that the historical context of the document greatly influences its interpretation by modern scholars. \\

\bottomrule
\end{tabular}
}
\end{table*}

\subsection{Hierarchical Multi-axis Taxonomy}
\label{sec:taxonomy}

MINT-Bench is designed to cover a broad range of control cases in instruction-following TTS, while keeping the benchmark space structured, extensible, and diagnostically meaningful. Rather than organizing the benchmark as a flat prompt inventory, we formulate it through a \emph{Hierarchical Multi-axis Taxonomy}, as illustrated in Figure~\ref{fig:overview}. The central design principle is that MINT-Bench is grounded in a compact set of atomic acoustic attributes, and richer benchmark cases are constructed by organizing and structurally realizing these attributes under different difficulty levels, control domains, control specifications, and fine-grained patterns.

The core of the taxonomy is ten atomic acoustic attributes, summarized in Table~\ref{tab:attributes}. These attributes form the primary timbre and style inventory of MINT-Bench. Specifically, the timbre-related attributes include \emph{age}, \emph{gender}, \emph{pitch}, and \emph{texture}, while the style-related attributes include \emph{speed}, \emph{volume}, \emph{emotion}, \emph{accent}, \emph{personality}, and \emph{tone}. These ten attributes serve not only as the benchmark's atomic control units, but also as the basic building blocks for more structured timbre- and style-oriented benchmark cases.

Importantly, MINT-Bench does not evaluate atomic attributes only in isolation. Rather, the benchmark primarily expands the same timbre/style inventory into richer control cases, such as simple and complex compositions, dynamic or layered realizations, and conflictive settings where multiple cues must be jointly resolved. Persona- and scenario-based prompts are treated in the same spirit: not as independent atomic attributes, but as higher-order composite realizations that implicitly bundle multiple timbre and style cues. In addition, MINT-Bench further includes a distinct \emph{special} extra-vocal branch covering abnormal vocalization and nonverbal-event control.

To make this organization explicit, the taxonomy is structured along four axes. The first axis is \emph{Difficulty Level}, which divides benchmark cases into \emph{Easy}, \emph{Hard}, and \emph{Special}. Easy cases mainly involve localized and relatively direct control over the atomic inventory, including single-attribute control and simple static combinations. Hard cases involve structurally richer realizations over the same inventory, such as complex composition, dynamic change, layered or conflictive control, and persona- or scenario-based prompting. Special cases cover extra-vocal control beyond ordinary timbre/style manipulation, including abnormal vocalization and nonverbal phenomena.

The second axis is \emph{Control Domain}, which specifies the functional domain of the target control case. In MINT-Bench, \emph{timbre} and \emph{style} denote atomic attribute domains grounded in the ten-attribute inventory, \emph{composition} denotes joint realization over multiple atomic timbre/style attributes, and \emph{extra-vocal} denotes the special case for abnormal or nonverbal vocal behavior. The third axis is \emph{Control Specification}, which describes how the target control is expressed in the instruction, for example, through tags, direct natural-language description, simple or complex compositional description, or implicit versus explicit extra-vocal cues. The fourth axis is \emph{Fine-grained Control Pattern}, which characterizes the structural form of the target case, such as static, dynamic, layered, conflictive, scenario-based, or character-based realization.

A \emph{taxonomy node} corresponds to one valid coordinate in this four-axis space and serves as the basic benchmark unit in MINT-Bench. Each node specifies a well-defined control case in terms of difficulty, domain, specification, and structural pattern, while leaving the final linguistic surface form open. In practice, taxonomy nodes are later instantiated through the data construction pipeline into concrete instruction-text pairs. Therefore, the taxonomy determines the benchmark's coverage and internal organization, while the final benchmark items remain natural-language realizations rather than fixed label strings.

\begin{figure*}[h]
  \centering
  \includegraphics[width=0.80\linewidth]{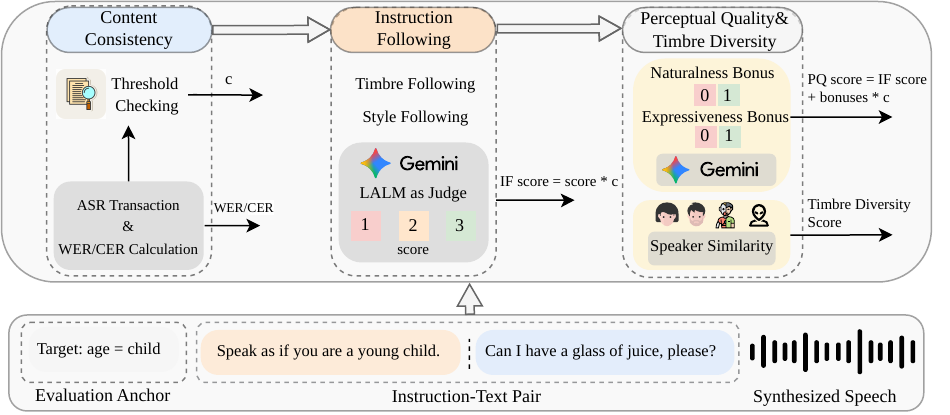}
    \caption{Hierarchical hybrid evaluation protocol of MINT-Bench. The pipeline progressively evaluates content consistency, instruction following, perceptual quality, and timbre diversity.
    }
  \label{fig:evaluation}
\end{figure*}

\subsection{Three-stage Data Construction}
\label{sec:data_construction}

Building on the taxonomy above, MINT-Bench constructs benchmark instances through a controllable multi-stage pipeline rather than generating final instruction-text pairs in a single step. This design is motivated by the heterogeneity of instruction-following TTS control cases. Direct one-shot generation is prone to attribute omission, uncontrolled semantic drift, leakage of target cues into the synthesis text, and internal inconsistency between the intended control and the final instruction. MINT-Bench, therefore, separates benchmark design from linguistic realization, allowing coverage, controllability, and naturalness to be balanced more effectively.

The construction pipeline consists of three stages: \emph{node specification}, \emph{structured label planning}, and \emph{instruction-text realization}. Representative examples are shown in Table~\ref{tab:data_construction_examples}.

\noindent\textbf{Stage 1: Taxonomy Node Specification.}
The first stage specifies valid benchmark cases under the taxonomy and prepares the internal inventory needed for controlled instantiation. For atomic control cases, this inventory is grounded in the ten foundational attributes described above. For higher-order cases, it additionally includes structured resources for composition, persona- or scenario-related labels, and extra-vocal control. The purpose of this stage is not to define a fixed external label vocabulary, but to provide internal support for coverage control, ambiguity reduction, and systematic case generation. Each taxonomy node is therefore associated with a specification that defines its \emph{Difficulty Level}, \emph{Control Domain}, \emph{Control Specification}, \emph{Fine-grained Control Pattern}, target language, and item budget.

\noindent\textbf{Stage 2: Structured Label Planning.}
The second stage converts each node specification into an intermediate structured plan. This plan serves as a normalized control scaffold between the taxonomy and the final natural-language benchmark item. It specifies the target values or control intent to be realized, together with semantic setup, text length, and realization constraints. This intermediate layer is essential for quality control. It reduces uncontrolled generation variance, prevents trivial target leakage into the synthesis text, and ensures that the final instance remains faithful to the intended control case while leaving sufficient flexibility for natural language realization. For simple atomic cases, the plan is usually compact and localized. While for compositional, persona-based, dynamic, or extra-vocal cases, the plan typically includes richer structural constraints so that the resulting instruction remains coherent and sufficiently informative.

\noindent\textbf{Stage 3: Instruction-Text Pair Construction.}
In the final stage, each structured plan is realized as a natural-language instruction paired with a synthesis text. This stage produces the benchmark items that are actually used to in TTS systems. Importantly, the final instruction is not restricted to a fixed label string. Depending on the control case, it may take the form of a short directive, a compositional description, or a paragraph-level prompt that integrates multiple vocal cues. The synthesis text is generated under constraints designed to minimize direct lexical leakage of the control target, so that the benchmark evaluates speech control rather than simple textual cue matching. The resulting instruction-text pairs therefore preserve the structural intent defined by the taxonomy while remaining natural and diverse in surface form.

Overall, the construction pipeline can be viewed as a progressive constrained-realization process. The taxonomy specifies what kinds of benchmark cases should be covered, node specification defines valid instances under that space, structured planning normalizes their control intent, and final realization converts them into natural-language benchmark items. This design makes MINT-Bench scalable, extensible, and substantially more controllable than direct end-to-end prompt generation.

\begin{table*}[t]
\centering
\small
\setlength{\tabcolsep}{3.8pt}
\renewcommand{\arraystretch}{1.1}

\newcommand{\scorecell}[1]{#1}
\newcommand{\bestscore}[1]{\textbf{#1}}
\newcommand{\secondscore}[1]{\underline{#1}}
\newcommand{\pairscore}[2]{\makecell[c]{#1 / #2}}

\caption{Detailed results on English (en) and Chinese (zh). For Overall, Timbre, and Style, scores are reported as \textit{Instruction-Following (IF) score / Perceptual Quality score}. Within each system category (Commercial and Open-source), the best results are \textbf{bolded} and the second-best are \underline{underlined}. Abbreviations include: Lang (Language), WER (Word Error Rate), TDS (Timbre Diversity Score), B-Comp. (Binary-attribute Composition), M-Comp. (Multi-attribute Composition), Disfl. (Disfluency), Dysph. (Dysphonia), Impl. (Implicit nonverbal), and Expl. (Explicit nonverbal).}
\label{tab:main_results_en_zh_full}
\resizebox{\textwidth}{!}{
\begin{tabular}{>{\raggedright\arraybackslash\bfseries\normalsize}l l *{18}{c}}
\toprule
\multirow{3}{*}{\normalsize\bfseries Model} 
& \multirow{3}{*}{\normalsize\bfseries Lang} 
& \multirow{3}{*}{\normalsize\bfseries WER $\downarrow$} 
& \multirow{3}{*}{\normalsize\bfseries TDS $\uparrow$} 
& \multirow{3}{*}{\normalsize\bfseries Overall $\uparrow$} 
& \multirow{3}{*}{\normalsize\bfseries Timbre $\uparrow$} 
& \multirow{3}{*}{\normalsize\bfseries Style $\uparrow$} 
& \multicolumn{3}{c}{\normalsize\bfseries Easy $\uparrow$} 
& \multicolumn{6}{c}{\normalsize\bfseries Hard $\uparrow$} 
& \multicolumn{4}{c}{\normalsize\bfseries Special $\uparrow$} \\
\cmidrule(lr){8-10} \cmidrule(lr){11-16} \cmidrule(lr){17-20}
& & & & & & 
& \footnotesize\bfseries Tag & \footnotesize\bfseries Direct & \footnotesize\bfseries B-Comp.
& \footnotesize\bfseries M-Comp. & \footnotesize\bfseries Dynamic & \footnotesize\bfseries Layered & \footnotesize\bfseries Conflict & \footnotesize\bfseries Scenario & \footnotesize\bfseries Character
& \footnotesize\bfseries Disfl. & \footnotesize\bfseries Dysph. & \footnotesize\bfseries Impl. & \footnotesize\bfseries Expl. \\
\midrule

\rowcolor{tableblue}
\multicolumn{20}{c}{\normalsize\bfseries Commercial Systems} \\
\midrule

\multirow{2}{*}{Gemini 2.5-Flash}
& en & 1.4\%
& \secondscore{0.46}
& \pairscore{\bestscore{2.44}}{\bestscore{3.66}}
& \pairscore{2.33}{\bestscore{3.37}}
& \pairscore{\bestscore{2.59}}{\bestscore{3.96}}
& \bestscore{3.74} & \bestscore{3.81} & \bestscore{4.07}
& \bestscore{3.64} & \secondscore{3.54} & \bestscore{3.94} & \scorecell{3.25} & \secondscore{3.38} & \secondscore{3.22}
& \bestscore{2.55} & \secondscore{4.08} & \bestscore{3.94} & \scorecell{3.56} \\
& zh & 1.6\%
& 0.32
& \pairscore{\secondscore{2.09}}{\bestscore{2.95}}
& \pairscore{2.09}{2.79}
& \pairscore{\secondscore{2.11}}{\secondscore{3.04}}
& \scorecell{2.85} & \secondscore{3.19} & \bestscore{3.52}
& \bestscore{3.05} & \bestscore{2.86} & \scorecell{2.22} & \scorecell{1.98} & \bestscore{2.56} & \bestscore{2.63}
& \secondscore{1.75} & \secondscore{4.12} & \bestscore{3.66} & \secondscore{3.41} \\
\midrule

\multirow{2}{*}{Gemini 2.5-Pro}
& en & 1.4\%
& \bestscore{0.57}
& \pairscore{\secondscore{2.39}}{\secondscore{3.45}}
& \pairscore{\secondscore{2.34}}{3.17}
& \pairscore{\secondscore{2.53}}{\secondscore{3.73}}
& \secondscore{3.49} & \secondscore{3.64} & \secondscore{3.60}
& \secondscore{3.32} & \bestscore{3.80} & \secondscore{3.38} & \bestscore{3.57} & \scorecell{3.26} & \scorecell{2.91}
& \secondscore{2.40} & \bestscore{4.21} & \secondscore{3.50} & \secondscore{3.58} \\
& zh & 1.8\%
& \bestscore{0.39}
& \pairscore{\bestscore{2.13}}{\secondscore{2.93}}
& \pairscore{\bestscore{2.20}}{\secondscore{2.85}}
& \pairscore{\bestscore{2.22}}{\bestscore{3.11}}
& \bestscore{2.91} & \scorecell{3.09} & \secondscore{3.43}
& \secondscore{2.93} & \secondscore{2.82} & \secondscore{2.46} & \bestscore{2.84} & \scorecell{2.12} & \scorecell{2.31}
& \bestscore{2.15} & \bestscore{4.21} & \secondscore{3.53} & \bestscore{3.62} \\
\midrule

\multirow{2}{*}{ElevenLabs-ttv-v3}
& en & 1.4\%
& 0.31
& \pairscore{2.23}{3.12}
& \pairscore{\bestscore{2.44}}{\secondscore{3.27}}
& \pairscore{2.25}{3.13}
& \scorecell{3.22} & \scorecell{3.06} & \secondscore{3.65}
& \scorecell{3.00} & \scorecell{2.50} & \scorecell{3.21} & \secondscore{3.30} & \bestscore{3.95} & \bestscore{3.66}
& \scorecell{0.89} & \scorecell{2.08} & \scorecell{1.61} & \scorecell{3.40} \\
& zh & 1.7\%
& 0.18
& \pairscore{1.98}{2.61}
& \pairscore{\secondscore{2.18}}{2.71}
& \pairscore{1.94}{2.56}
& \scorecell{2.65} & \scorecell{2.39} & \scorecell{2.90}
& \scorecell{2.96} & \scorecell{1.67} & \bestscore{2.56} & \secondscore{2.49} & \secondscore{2.56} & \bestscore{3.06}
& \scorecell{1.10} & \scorecell{3.33} & \scorecell{1.49} & \scorecell{3.26} \\
\midrule

\multirow{2}{*}{MiniMax-Speech-2.7}
& en & 1.4\%
& 0.44
& \pairscore{1.95}{2.62}
& \pairscore{2.12}{2.82}
& \pairscore{2.04}{2.77}
& \scorecell{2.85} & \scorecell{2.65} & \scorecell{3.20}
& \scorecell{3.02} & \scorecell{1.99} & \scorecell{2.02} & \scorecell{2.83} & \scorecell{3.00} & \scorecell{2.93}
& \scorecell{1.00} & \scorecell{1.17} & \scorecell{1.48} & \scorecell{1.00} \\
& zh & 1.4\%
& \secondscore{0.38}
& \pairscore{1.95}{2.62}
& \pairscore{2.17}{\bestscore{2.88}}
& \pairscore{2.07}{2.86}
& \bestscore{2.84} & \scorecell{2.86} & \scorecell{3.47}
& \scorecell{2.99} & \scorecell{2.20} & \secondscore{2.52} & \scorecell{2.38} & \bestscore{2.80} & \scorecell{2.25}
& \scorecell{1.05} & \scorecell{2.50} & \scorecell{1.44} & \scorecell{1.12} \\
\midrule

Hume-Octave1
& en & 1.6\%
& 0.37
& \pairscore{1.86}{2.51}
& \pairscore{2.08}{2.86}
& \pairscore{1.95}{2.69}
& \scorecell{3.00} & \scorecell{2.98} & \scorecell{3.11}
& \scorecell{2.71} & \scorecell{1.99} & \scorecell{1.62} & \scorecell{1.68} & \scorecell{2.56} & \scorecell{2.28}
& \scorecell{1.00} & \scorecell{1.00} & \scorecell{1.00} & \scorecell{1.03} \\
\midrule

\multirow{2}{*}{GPT-4o-Mini-TTS}
& en & 1.7\%
& 0.37
& \pairscore{1.69}{2.15}
& \pairscore{1.70}{2.09}
& \pairscore{1.78}{2.30}
& \scorecell{2.07} & \scorecell{2.42} & \scorecell{2.16}
& \scorecell{2.22} & \scorecell{2.24} & \scorecell{2.19} & \scorecell{2.40} & \scorecell{1.89} & \scorecell{1.79}
& \scorecell{1.30} & \scorecell{2.71} & \scorecell{2.19} & \scorecell{2.03} \\
& zh & 1.7\%
& 0.25
& \pairscore{1.56}{1.82}
& \pairscore{1.58}{1.72}
& \pairscore{1.63}{1.92}
& \scorecell{1.75} & \scorecell{2.01} & \scorecell{2.06}
& \scorecell{1.86} & \scorecell{2.25} & \scorecell{1.23} & \scorecell{1.68} & \scorecell{1.31} & \scorecell{1.54}
& \scorecell{1.11} & \scorecell{2.71} & \scorecell{2.06} & \scorecell{1.91} \\

\midrule
\rowcolor{tableblue}
\multicolumn{20}{c}{\normalsize\bfseries Open-source Systems} \\
\midrule

\multirow{2}{*}{Qwen3TTS-12Hz-1.7B-VD}
& en & 1.5\%
& 0.37
& \pairscore{\bestscore{2.12}}{\bestscore{2.94}}
& \pairscore{\bestscore{2.28}}{\bestscore{3.16}}
& \pairscore{\bestscore{2.16}}{\bestscore{3.03}}
& \bestscore{3.14} & \bestscore{2.96} & \bestscore{3.44}
& \bestscore{3.27} & \scorecell{1.99} & \bestscore{2.90} & \scorecell{2.79} & \bestscore{2.97} & \bestscore{2.96}
& \secondscore{1.20} & \bestscore{3.29} & \secondscore{2.44} & \bestscore{2.03} \\
& zh & 1.1\%
& 0.28
& \pairscore{\bestscore{2.16}}{\bestscore{3.12}}
& \pairscore{\bestscore{2.29}}{\bestscore{3.30}}
& \pairscore{\bestscore{2.23}}{\bestscore{3.27}}
& \bestscore{3.35} & \bestscore{3.38} & \bestscore{4.05}
& \bestscore{3.26} & \bestscore{2.06} & \bestscore{2.69} & \scorecell{2.67} & \bestscore{2.63} & \bestscore{2.66}
& \scorecell{1.40} & \bestscore{4.25} & \bestscore{2.97} & \bestscore{2.34} \\
\midrule

\multirow{2}{*}{MOSS-VoiceGenerator}
& en & 2.8\%
& \secondscore{0.47}
& \pairscore{\secondscore{1.90}}{\secondscore{2.53}}
& \pairscore{\secondscore{2.01}}{\secondscore{2.61}}
& \pairscore{\secondscore{1.99}}{\secondscore{2.72}}
& \secondscore{2.52} & \secondscore{2.57} & \secondscore{3.09}
& \secondscore{2.78} & \bestscore{2.35} & \secondscore{2.64} & \bestscore{2.84} & \scorecell{2.17} & \secondscore{2.28}
& \bestscore{1.50} & \secondscore{2.92} & \bestscore{2.77} & \secondscore{1.28} \\
& zh & 3.1\%
& \secondscore{0.35}
& \pairscore{1.93}{2.63}
& \pairscore{1.98}{2.60}
& \pairscore{2.00}{2.79}
& \secondscore{2.70} & \secondscore{2.66} & \secondscore{2.90}
& \secondscore{3.15} & \scorecell{2.01} & \secondscore{2.56} & \bestscore{2.77} & \scorecell{2.10} & \secondscore{2.41}
& \secondscore{1.50} & \secondscore{3.67} & \secondscore{2.70} & \bestscore{2.34} \\
\midrule

\multirow{2}{*}{MiMo-Audio-7B-Instruct}
& en & 3.5\%
& \bestscore{0.52}
& \pairscore{1.75}{2.22}
& \pairscore{1.80}{2.17}
& \pairscore{1.83}{2.37}
& \scorecell{2.42} & \scorecell{2.28} & \scorecell{2.49}
& \scorecell{2.49} & \secondscore{1.81} & \scorecell{1.46} & \secondscore{2.25} & \secondscore{2.53} & \scorecell{2.13}
& \scorecell{1.05} & \scorecell{2.54} & \scorecell{1.91} & \scorecell{1.03} \\
& zh & 3.5\%
& \bestscore{0.37}
& \pairscore{1.75}{2.22}
& \pairscore{1.78}{2.17}
& \pairscore{1.83}{2.35}
& \scorecell{2.37} & \scorecell{2.25} & \scorecell{2.80}
& \scorecell{2.15} & \secondscore{1.97} & \scorecell{1.46} & \secondscore{2.11} & \secondscore{2.43} & \scorecell{1.90}
& \bestscore{1.65} & \scorecell{3.21} & \scorecell{2.34} & \scorecell{1.41} \\
\midrule

\multirow{2}{*}{Ming-omni-tts-16.8B-A3B}
& en & 1.8\%
& 0.35
& \pairscore{1.51}{1.79}
& \pairscore{1.68}{1.94}
& \pairscore{1.58}{1.89}
& \scorecell{2.17} & \scorecell{1.87} & \scorecell{2.43}
& \scorecell{1.52} & \scorecell{1.92} & \scorecell{1.62} & \scorecell{1.38} & \scorecell{1.99} & \scorecell{1.55}
& \scorecell{1.00} & \scorecell{1.00} & \scorecell{1.03} & \scorecell{1.25} \\
& zh & 0.7\%
& 0.22
& \pairscore{\secondscore{1.95}}{\secondscore{2.68}}
& \pairscore{\secondscore{2.10}}{\secondscore{2.85}}
& \pairscore{\secondscore{2.06}}{\secondscore{2.92}}
& \scorecell{2.99} & \scorecell{2.84} & \scorecell{3.66}
& \scorecell{2.91} & \scorecell{2.44} & \scorecell{2.40} & \scorecell{2.13} & \scorecell{2.44} & \scorecell{2.42}
& \scorecell{1.00} & \scorecell{2.33} & \scorecell{2.22} & \scorecell{1.12} \\
\midrule

\multirow{2}{*}{Ming-omni-tts-0.5B}
& en & 2.4\%
& 0.35
& \pairscore{1.41}{1.70}
& \pairscore{1.55}{1.91}
& \pairscore{1.46}{1.75}
& \scorecell{1.85} & \scorecell{1.88} & \scorecell{2.08}
& \scorecell{1.48} & \scorecell{1.55} & \scorecell{1.12} & \scorecell{1.29} & \scorecell{1.73} & \scorecell{1.54}
& \scorecell{1.00} & \scorecell{1.04} & \scorecell{1.13} & \scorecell{1.00} \\
& zh & 0.6\%
& 0.23
& \pairscore{1.87}{2.56}
& \pairscore{2.04}{2.81}
& \pairscore{2.01}{2.79}
& \scorecell{2.80} & \scorecell{2.76} & \scorecell{3.45}
& \scorecell{2.98} & \scorecell{2.07} & \scorecell{2.08} & \scorecell{2.14} & \scorecell{2.16} & \scorecell{2.47}
& \scorecell{1.00} & \scorecell{2.00} & \scorecell{1.68} & \scorecell{1.13} \\
\midrule

Parler-TTS Large
& en & 4.4\%
& 0.42
& \pairscore{1.38}{1.52}
& \pairscore{1.61}{1.71}
& \pairscore{1.49}{1.68}
& \scorecell{1.61} & \scorecell{1.85} & \scorecell{1.88}
& \scorecell{1.76} & \scorecell{0.88} & \scorecell{1.55} & \scorecell{1.39} & \scorecell{1.03} & \scorecell{1.01}
& \scorecell{1.15} & \scorecell{2.29} & \scorecell{1.12} & \scorecell{1.00} \\
\midrule

Parler-TTS Mini
& en & 17.5\%
& 0.33
& \pairscore{1.04}{1.13}
& \pairscore{1.11}{1.17}
& \pairscore{1.13}{1.25}
& \scorecell{1.29} & \scorecell{1.39} & \scorecell{1.04}
& \scorecell{1.40} & \scorecell{0.82} & \scorecell{1.01} & \scorecell{0.80} & \scorecell{0.48} & \scorecell{0.54}
& \scorecell{1.00} & \scorecell{3.04} & \scorecell{1.25} & \scorecell{1.00} \\
\midrule

VoiceSculptor
& zh & 9.2\%
& 0.23
& \pairscore{1.59}{2.01}
& \pairscore{1.55}{1.84}
& \pairscore{1.62}{2.09}
& \scorecell{2.13} & \scorecell{1.80} & \scorecell{2.33}
& \scorecell{2.08} & \scorecell{1.82} & \scorecell{2.02} & \scorecell{1.54} & \scorecell{1.78} & \scorecell{2.00}
& \scorecell{1.35} & \scorecell{3.46} & \scorecell{2.38} & \scorecell{1.16} \\

\bottomrule
\end{tabular}
}
\end{table*}

\subsection{Hierarchical Hybrid Evaluation}
\label{sec:evaluation_methodology}

Given an instruction-text pair and the synthesized speech produced by a candidate TTS system, MINT-Bench applies a hierarchical hybrid evaluation protocol, as illustrated in Figure~\ref{fig:evaluation}. The pipeline evaluates generated speech from three complementary perspectives: \emph{content consistency}, \emph{instruction following}, and \emph{perceptual quality}. This design reflects the fact that instruction-following TTS should not be judged by a single dimension: a successful system should preserve the intended linguistic content, realize the requested control target, and deliver the result with convincing perceptual quality.

\noindent\textbf{Stage 1: Content Consistency.}
We first assess whether the synthesized speech preserves the intended linguistic content. This stage uses ASR transcription together with sample-level WER checking. This WER-based content consistency check is applied only to subsets for which ASR transcription remains a reasonable proxy for content preservation. It is not applied to accent-related subsets, because ASR models are often unstable under strong accent variation, nor to \textit{Special} subsets, because extra-vocal phenomena such as disfluencies, dysphonia, or nonverbal events can substantially distort ASR output even when the intended control is correctly realized. In these cases, WER is not a reliable indicator of whether the content target has been preserved.

We further define a WER threshold for each language. Specifically, for a given language, we collect the overall WER values of the system, remove the highest and lowest values, and compute the mean of the remaining systems. This trimmed mean serves as the language-specific WER threshold, denoted by $\tau_{\ell}$ for language $\ell$. For each sample, if its WER is below $\tau_{\ell}$, we regard it as a \emph{low-WER} sample; otherwise, it is regarded as a \emph{high-WER} sample.

Based on the threshold above, we define the content consistency coefficient as the proportion of low-WER samples:

\begin{equation}
p = \frac{m_{\text{low}}}{N_{\text{cc}}},
\end{equation}
where $m_{\text{low}}$ is the number of evaluated samples whose WER falls below the corresponding language-specific threshold, and $N_{\text{cc}}$ is the number of samples for which the content consistency check is applicable. 

Under this formulation, Stage~1 does not discard samples from later analysis; instead, it penalizes systems that frequently produce content deviations by scaling the output scores of both Stage~2 (instruction following) and Stage~3 (perceptual quality) with the content consistency coefficient $c$.

\noindent\textbf{Stage 2: Instruction Following.}
MINT-Bench next evaluates whether the synthesized speech follows the intended control case. This stage is performed by a Large Audio Language Model (LALM), i.e., Gemini, using task-aware judging prompts derived from the taxonomy. Since \textit{Easy}, \textit{Hard}, and \textit{Special} cases require different kinds of evidence, we use three prompt families for judging, one for each task stratum, with case-specific definitions and constraints embedded in each family.

\begin{table*}[h]
\centering
\footnotesize 
\setlength{\tabcolsep}{3.5pt}
\renewcommand{\arraystretch}{1.1} 

\newcommand{\bestscore}[1]{\textbf{#1}}
\newcommand{\secondscore}[1]{\underline{#1}}
\newcommand{\pairscore}[2]{\makecell[c]{#1 / #2}}
\caption{Multilingual evaluation results across ten languages. The columns of Overall, Timbre, and Style are reported as \textit{Instruction-Following (IF) / Perceptual Quality (PE) scores}, while the per-language columns report \textit{PE score}. Within each system category (Commercial and Open-source), the best results are \textbf{bolded} and the second-best are \underline{underlined}. }
\label{tab:multilingual_leaderboard_full}

\resizebox{\textwidth}{!}{
\begin{tabular}{l ccc cccccccccc} 
\toprule
\bfseries Model 
& \bfseries \makecell[c]{Overall  $\uparrow$} 
& \bfseries \makecell[c]{Timbre  $\uparrow$} 
& \bfseries \makecell[c]{Style  $\uparrow$}
& \bfseries Chinese $\uparrow$ & \bfseries English $\uparrow$ & \bfseries German $\uparrow$ & \bfseries Spanish $\uparrow$ & \bfseries French $\uparrow$ & \bfseries Italian $\uparrow$ & \bfseries Japanese $\uparrow$ & \bfseries Korean $\uparrow$ & \bfseries Portuguese $\uparrow$ & \bfseries Russian $\uparrow$ \\
\midrule

\rowcolor{tableblue}
\multicolumn{14}{c}{\scriptsize\bfseries Commercial Systems} \\
\midrule

Gemini 2.5-Flash
& \pairscore{\bestscore{2.50}}{\bestscore{3.75}}
& \pairscore{2.47}{\bestscore{3.58}}
& \pairscore{\bestscore{2.60}}{\bestscore{3.95}}
& \bestscore{2.95} & \bestscore{3.66} & \bestscore{3.96} & \bestscore{3.70} & \bestscore{3.70} & \bestscore{3.72} & \secondscore{4.03} & \secondscore{4.08} & \bestscore{3.80} & \bestscore{3.94} \\

Gemini 2.5-Pro
& \pairscore{\secondscore{2.48}}{\secondscore{3.63}}
& \pairscore{\secondscore{2.50}}{\secondscore{3.53}}
& \pairscore{\secondscore{2.58}}{\secondscore{3.82}}
& \secondscore{2.93} & \secondscore{3.45} & \secondscore{3.72} & \secondscore{3.49} & \secondscore{3.64} & \secondscore{3.51} & \bestscore{4.05} & \bestscore{4.13} & \secondscore{3.64} & \secondscore{3.80} \\

ElevenLabs-ttv-v3
& \pairscore{2.31}{3.22}
& \pairscore{\bestscore{2.55}}{3.41}
& \pairscore{2.28}{3.18}
& 2.61 & 3.12 & 3.52 & 3.25 & 3.28 & 3.00 & 3.39 & 3.46 & 3.24 & 3.29 \\

MiniMax-Speech-2.7
& \pairscore{2.03}{2.70}
& \pairscore{2.26}{2.99}
& \pairscore{2.13}{2.91}
& 2.62 & 2.62 & 2.74 & 2.87 & 2.73 & 2.71 & 2.14 & 3.07 & 2.94 & 2.61 \\

GPT-4o-Mini-TTS
& \pairscore{1.67}{1.99}
& \pairscore{1.78}{2.03}
& \pairscore{1.73}{2.11}
& 1.82 & 2.15 & 1.77 & 1.68 & 1.77 & 1.60 & 2.30 & 2.33 & 2.15 & 2.35 \\

\midrule
\rowcolor{tableblue}
\multicolumn{14}{c}{\scriptsize\bfseries Open-source Systems} \\
\midrule

Qwen3TTS-12Hz-1.7B-VD
& \pairscore{\bestscore{2.29}}{\bestscore{3.19}}
& \pairscore{\bestscore{2.46}}{\bestscore{3.36}}
& \pairscore{\bestscore{2.34}}{\bestscore{3.35}}
& \bestscore{3.12} & \bestscore{2.94} & \bestscore{3.57} & \bestscore{3.29} & \bestscore{3.17} & \bestscore{2.94} & \bestscore{3.32} & \bestscore{3.77} & \bestscore{3.21} & \bestscore{2.63} \\

Ming-omni-tts-16.8B-A3B
& \pairscore{\secondscore{1.13}}{\secondscore{1.25}}
& \pairscore{\secondscore{1.27}}{\secondscore{1.38}}
& \pairscore{\secondscore{1.11}}{\secondscore{1.24}}
& \secondscore{2.68} & 1.79 & \secondscore{1.39} & \secondscore{1.30} & \secondscore{1.26} & \secondscore{1.21} & \secondscore{0.55} & 0.25 & \secondscore{1.10} & \secondscore{0.91} \\

MOSS-VoiceGenerator
& \pairscore{0.87}{1.04}
& \pairscore{0.79}{0.93}
& \pairscore{0.78}{0.97}
& 2.63 & \secondscore{2.53} & 0.67 & 1.20 & 0.60 & 0.89 & 0.39 & \secondscore{0.58} & 0.65 & 0.31 \\

Mimo-Audio-7B-Instruct
& \pairscore{0.62}{0.73}
& \pairscore{0.49}{0.56}
& \pairscore{0.49}{0.60}
& 2.22 & 2.22 & 0.28 & 0.35 & 0.32 & 0.35 & 0.59 & 0.49 & 0.24 & 0.19 \\

Ming-omni-tts-0.5B
& \pairscore{0.54}{0.64}
& \pairscore{0.51}{0.62}
& \pairscore{0.43}{0.54}
& 2.56 & 1.70 & 0.26 & 0.31 & 0.39 & 0.39 & 0.16 & 0.16 & 0.29 & 0.17 \\

\bottomrule
\end{tabular}
}
\end{table*}

For each sample, the LALM assigns a discrete instruction\-/following score $s_i \in \{1,2,3\}$, corresponding to poor, partial, or strong realization of the intended control target. This score reflects how well the synthesized speech satisfies the control requirement expressed by the instruction, regardless of whether the target mainly concerns timbre, style, composition, or extra-vocal behavior. The system-level instruction-following score (IF Score) is then computed as
\begin{equation}
IF\ Score =
p \cdot \frac{1}{N}\sum_{i=1}^{N} s_i,
\end{equation}
where $c$ is the content consistency coefficient from Stage~1 and $N$ is the total number of evaluated samples in this stage. In this way, systems that frequently violate content consistency are penalized even if their raw instruction\-/following scores are high. For analysis purposes, these sample-level scores can later be grouped and summarized according to the control attributes involved in the corresponding instructions.

\noindent\textbf{Stage 3: Perceptual Quality and Timbre Diversity.}
Accurate instruction following does not by itself guarantee high-quality speech. A system may realize the requested control target while still sounding unnatural, dull, or weakly expressive. MINT-Bench therefore introduces a conditional \emph{perceptual quality} stage that is applied only to samples with strong instruction-following performance, namely samples with $s_i = 3$ from Stage~2.

This stage contains two binary perceptual bonuses: naturalness $B_{\text{nat},i} \in \{0,1\}$ and expressiveness $B_{\text{exp},i} \in \{0,1\}$. These bonuses are assessed only for samples that already show strong realization of the intended control target. For all other samples, both bonuses are set to 0 by definition. Under this design, Stage~3 refines the upper range of system performance without changing the ranking structure among clearly failed or partially realized samples.

\begin{table*}[h]
\centering
\small
\setlength{\tabcolsep}{4.5pt}
\renewcommand{\arraystretch}{1.2}

\caption{Correlation of instruction-following evaluation between human and LALM. Blue values denote standard deviations.}
\label{tab:human_model_correlation}
\resizebox{\textwidth}{!}{
\begin{tabular}{l cccccccccc}
\toprule
\bfseries Metric
& \bfseries Chinese & \bfseries English & \bfseries German & \bfseries Spanish & \bfseries French & \bfseries Italian & \bfseries Japanese & \bfseries Korean & \bfseries Portuguese & \bfseries Russian \\
\midrule

Inter-Human Agreement
& 69.45 {\scriptsize$\pm$ \textcolor{blue!70!black}{0.79}}
& 77.32 {\scriptsize$\pm$ \textcolor{blue!70!black}{0.72}}
& 74.55 {\scriptsize$\pm$ \textcolor{blue!70!black}{1.34}}
& 75.20 {\scriptsize$\pm$ \textcolor{blue!70!black}{1.41}}
& 75.88 {\scriptsize$\pm$ \textcolor{blue!70!black}{1.29}}
& 74.90 {\scriptsize$\pm$ \textcolor{blue!70!black}{1.38}}
& 79.12 {\scriptsize$\pm$ \textcolor{blue!70!black}{1.22}}
& 78.40 {\scriptsize$\pm$ \textcolor{blue!70!black}{1.26}}
& 75.10 {\scriptsize$\pm$ \textcolor{blue!70!black}{1.46}}
& 74.10 {\scriptsize$\pm$ \textcolor{blue!70!black}{1.52}} \\

Model--Consensus Human
& 67.12 {\scriptsize$\pm$ \textcolor{blue!70!black}{0.93}}
& 74.81 {\scriptsize$\pm$ \textcolor{blue!70!black}{0.86}}
& 72.18 {\scriptsize$\pm$ \textcolor{blue!70!black}{1.51}}
& 72.80 {\scriptsize$\pm$ \textcolor{blue!70!black}{1.57}}
& 73.42 {\scriptsize$\pm$ \textcolor{blue!70!black}{1.45}}
& 72.10 {\scriptsize$\pm$ \textcolor{blue!70!black}{1.54}}
& 77.35 {\scriptsize$\pm$ \textcolor{blue!70!black}{1.33}}
& 76.20 {\scriptsize$\pm$ \textcolor{blue!70!black}{1.37}}
& 72.50 {\scriptsize$\pm$ \textcolor{blue!70!black}{1.60}}
& 71.30 {\scriptsize$\pm$ \textcolor{blue!70!black}{1.66}} \\

\bottomrule
\end{tabular}
}
\end{table*}

The integrated system score is computed as
\begin{equation}
PE\ Score =
{IF\ Score}_i + p \cdot \frac{1}{N}\sum_{i=1}^{N}\left( B_{\text{nat},i} + B_{\text{exp},i}\right),
\end{equation}
where $c$ is the content consistency coefficient from Stage~1, $N$ is the total number of evaluated samples, and $s_i$ is the Stage~2 instruction-following score for sample $i$. Since the perceptual bonuses are only available when $s_i = 3$, the Stage~3 refinement raises a high-quality instruction-following sample from 3 to 4 or 5, depending on whether it receives one or both perceptual bonuses.

In addition, Stage~3 also evaluates diversity along the timbre dimension on selected subsets. For each selected instruction-text condition, the system generates multiple speech candidates and retains only those candidates that satisfy the Stage~2 requirement of strong instruction following. Let $m$ denote the number of retained candidates, and let $s_{ij}$ denote the speaker similarity between retained candidates $i$ and $j$. When $m < 2$, the timbre diversity score is set to 0. Otherwise, we compute the average pairwise similarity (APS):
\begin{equation}
\mathrm{APS} = \frac{2}{m(m-1)}\sum_{i<j}s_{ij},
\end{equation}
and define timbre diversity score (TDS) as
\begin{equation}
\mathrm{TDS} = (1 - \mathrm{APS}) \cdot \frac{m}{N},
\end{equation}

where $N$ is the total number of generated candidates for the same instruction-text condition. This formulation rewards systems that can produce multiple valid realizations with diverse speaker characteristics under the same control target. In this sense, timbre diversity complements the other evaluations by measuring one-to-many generation ability under a fixed instruction condition.

\section{Experiment}
\label{sec:experiment}

\subsection{Experimental Setup}
\label{sec:experiment_setup}

We evaluate MINT-Bench on a diverse set of both open-source and commercial instruction-following TTS systems. 
The open-source models include VoiceSculptor ~\cite{hu2026voicesculptorvoicedesigned} , MiMo-Audio-7B-Instruct ~\cite{mimo} , Qwen3TTS-12Hz-1.7B-VD ~\cite{qwen3tts} , Ming-omni-tts-16.8B-A3B, Ming-omni-tts-0.5B \footnote{\url{https://github.com/inclusionAI/Ming-omni-tts}}, Parler-TTS Mini, Parler-TTS Large ~\cite{parler} , and MOSS-VoiceGenerator ~\cite{huang2026mossvoicegeneratorcreaterealisticvoices} . 
The commercial systems include Gemini 2.5-Flash, Gemini 2.5-Pro ~\cite{gemini2.5} , GPT-4o-mini-TTS \footnote{\url{https://developers.openai.com/api/docs/models/gpt-4o-mini-tts}}, ElevenLabs-ttv-v3 \footnote{\url{https://elevenlabs.io/app/api/eleven_ttv_v3}}, MiniMax-Speech-2.7 \footnote{\url{https://api.minimaxi.com/v1/voice_design}}, and Hume-Octave1 \footnote{\url{https://www.hume.ai/}}. 
For each model, we evaluate the subset of benchmark languages that are supported by the model. 

MINT-Bench covers ten languages: Chinese, English, German, Italian, Portuguese, Spanish, Japanese, Korean, French, and Russian. 
Among them, Chinese and English serve as the primary evaluation languages. 
For these two languages, we construct large test splits following the full taxonomy-node specification, with approximately 1,000 instruction--text pairs per language. 
For the remaining languages, we construct mini splits derived from the same taxonomy configuration, each containing approximately 300 instruction-text pairs. 
We also prepare mini subsets for Chinese and English by sampling 30\% of the full node configuration for further investigation, so that the benchmark supports both full-scale and lower-cost evaluation under a shared taxonomy space.

For benchmark construction, we use Gemini~2.5~Pro as the LLM in the data construction pipeline. 
For evaluation, we use Gemini~3.1~Pro Preview as the primary LALM judge, following the hierarchical protocol described in Section~\ref{sec:evaluation_methodology}. 
In our experiments, the evaluation prompt is adopted according to the taxonomy prefix, with separate task-aware prompts for \textit{Easy}, \textit{Hard}, and \textit{Special} cases.
For content consistency, we use Paraformer-zh ~\cite{paraformer} for Chinese and Whisper Large-v3 ~\cite{whisper} for the other evaluated languages. 
For timbre diversity evaluation, we extract speaker embeddings with WavLM-Large ~\cite{wavlm} and compute similarity-based diversity scores over valid instruction-following samples. 
All reported metrics follow the unified hierarchical protocol introduced in Section~\ref{sec:evaluation_methodology}.

\subsection{Detailed Results on English and Chinese}
\label{sec:results_and_analysis}

Table~\ref{tab:main_results_en_zh_full} reports the detailed results on the English (en) and Chinese (zh) large splits. In accordance with the hierarchical multi-axis taxonomy, the evaluated subsets are organized into three difficulty strata, namely \textit{Easy}, \textit{Hard}, and \textit{Special}. 

The aggregated results \textit{Overall}, \textit{Timbre}, and \textit{Style} are reported as IF/PE score pairs, where IF denotes the Stage-2 instruction-following score and PE denotes the Stage-3 perceptual quality score. WER is reported as a content-consistency diagnostic, whereas TDS is reported as a timbre-diversity diagnostic under fixed instruction conditions.

\noindent\textbf{Bilingual overall results show a strong commercial frontier on English, but a much narrower gap on Chinese.}
Similar WER values do not necessarily translate into similar IF or PE scores, indicating that content consistency alone is insufficient to characterize instruction-following TTS. On the English large split, commercial systems define the leading frontier, with Gemini 2.5-Flash achieving the highest overall PE score (3.66), followed by Gemini 2.5-Pro (3.45). On Chinese, however, the gap narrows substantially: Qwen3TTS-12Hz-1.7B-VD attains the strongest overall PE score (3.12), slightly surpassing Gemini 2.5-Flash (2.95) and Gemini 2.5-Pro (2.93). This contrast suggests that leading open-source systems can already become highly competitive, and even take the lead, in localized bilingual settings.

\noindent\textbf{Timbre, style, and diversity reveal complementary aspects of controllability.}
The timbre and style aggregates show that strong overall performance does not imply uniform strength across attribute dimensions. Among commercial systems, Gemini 2.5-Flash is strongest on English timbre and style, while Gemini 2.5-Pro remains competitive and achieves the highest TDS in both English and Chinese within the commercial group. Among open-source systems, Qwen3TTS-12Hz-1.7B-VD consistently provides the strongest timbre and style control across both languages. At the same time, TDS does not simply track the main ranking: MiMo-Audio-7B-Instruct, for example, reaches relatively high TDS despite much weaker overall IF and PE scores. This confirms that timbre diversity should be interpreted as a separate conditional diagnostic of one-to-many generation ability, rather than as a surrogate for overall controllability.

\noindent\textbf{Difficulty-sensitive results expose a clear drop from easy control to structured and extra-vocal cases.}
Across both languages, the strongest systems handle many \textit{Easy} cases reliably, but performance declines once the control target becomes structurally richer. The \textit{Hard} subsets, especially \textit{Dynamic}, \textit{Layered}, \textit{Conflictive}, \textit{Scenario-based}, and \textit{Character-based} control, remain substantially more challenging, indicating that jointly resolving multiple interacting cues over the same timbre/style inventory is still far from solved. The \textit{Special} branch is the most difficult region overall, but its internal difficulty is not uniform: \textit{Disfluency Control} is consistently the hardest subset, whereas \textit{Dysphonia Control} and implicit or explicit extra-vocal control remain difficult but still partly reachable for the strongest systems. These patterns highlight the diagnostic value of MINT-Bench beyond a single aggregate ranking.

\subsection{Multilingual Results Analysis}

Table~\ref{tab:multilingual_leaderboard_full} summarizes multilingual evaluation results across ten languages. The summary columns, namely \textit{Overall}, \textit{Timbre}, and \textit{Style}, are reported as IF/PE score pairs. By contrast, the language-specific columns report PE scores only due to space limitations. 

\noindent\textbf{Multilingual results are led by a small frontier group, while multilingual robustness remains uneven.}
At the aggregate level, Gemini 2.5-Flash achieves the strongest PE scores on \textit{Overall}, \textit{Timbre}, and \textit{Style}, with Gemini 2.5-Pro ranking second. Among open-source systems, Qwen3TTS-12Hz-1.7B-VD is clearly the strongest multilingual baseline, leading the open-source group on all three summary aggregates. These results show that multilingual instruction-following TTS is still dominated by a small set of frontier systems, with a substantial gap between the strongest models and the remaining baselines.

\noindent\textbf{Language-wise results further show that strong multilingual averages do not imply uniform multilingual stability.}
Although Gemini 2.5-Flash is the strongest system overall, Gemini 2.5-Pro achieves higher PE scores on some individual languages, most notably Japanese and Korean, indicating that even top-tier systems exhibit language-specific capability peaks rather than flat multilingual profiles. Within the open-source group, Qwen3TTS-12Hz-1.7B-VD not only dominates the summary aggregates but also remains competitive on multiple individual languages, most notably achieving the strongest Chinese result in the table. By contrast, the remaining open-source systems show much weaker multilingual robustness, with performance often dropping sharply outside a limited subset of languages. Overall, Table~\ref{tab:multilingual_leaderboard_full} provides a diagnostic view of multilingual generalization by distinguishing broadly strong systems from those whose performance is driven by only a few localized strengths.

\subsection{Evaluation Consistency Analysis}
\label{sec:consistency_analysis}

To assess the reliability of the LALM-based judging pipeline, we conduct a multilingual human study covering all ten languages in MINT-Bench. For the English and Chinese large sets, we recruit 30 native speakers per language; for the remaining multilingual mini sets, we recruit 10 native speakers per language. The evaluation includes four representative systems, namely two commercial systems (Gemini 2.5-Flash and MiniMax-Speech-2.7) and two open-source systems (Qwen3TTS-12Hz-1.7B-VD and MiMo-Audio-7B-Instruct). To reduce the influence of severe content corruption on instruction-following judgments, the agreement analysis is restricted to one-fifth of the samples that pass the WER-based content check.

Table~\ref{tab:human_model_correlation} reports Spearman correlations for instruction-following judgments. \textbf{Inter-Human Agreement} denotes the average correlation among human annotators, while \textbf{Model--Consensus Human} denotes the correlation between the LALM-based judge and the mean human score. Across all languages, inter-human agreement ranges from 69.45 to 79.12, while model-consensus human agreement ranges from 67.12 to 77.35. The gap between the two remains consistently small, indicating that the automatic judge captures a stable shared standard and approaches the agreement level observed among human annotators. This trend is observed not only in the primary evaluation languages, English and Chinese, but also across the remaining multilingual settings, supporting the use of LALM-based judging as a scalable proxy for human assessment. At the same time, the model-based judge remains slightly below the inter-human ceiling, which is a more appropriate role for automatic evaluation than replacing human judgment entirely. The smaller standard deviations on English and Chinese are also consistent with the larger annotation pools used for these two primary languages.

\section{Conclusion}

We introduced \textbf{MINT-Bench}, a structured multilingual benchmark for instruction-following TTS that supports both systematic comparison and fine-grained diagnostic evaluation across diverse control settings and languages. By combining a hierarchical multi-axis taxonomy, a scalable construction pipeline, and a hierarchical hybrid evaluation protocol, MINT-Bench enables systematic and diagnostically meaningful assessment of controllable speech generation.
Experiments across ten languages show that instruction-following TTS remains far from solved. Strong content preservation does not necessarily imply strong controllability; frontier commercial systems still lead overall multilingual performance, while leading open-source systems can already be competitive in localized settings such as Chinese. The results further show that structurally richer controls, especially compositional and extra-vocal cases, remain major bottlenecks. Together with the strong agreement between LALM-based judging and human consensus, these findings suggest that MINT-Bench serves not only as a benchmark for system comparison but also as a diagnostic framework for understanding capability boundaries and guiding future research on controllable and multilingual speech generation.

\section{Limitations and Future Work}
Although MINT-Bench provides a structured and extensible benchmark for instruction-following TTS, several limitations remain.

First, MINT-Bench covers a bounded benchmark space rather than the full range of real-world instructions. More open-ended settings, such as long-form generation, dialogue interaction, richer discourse context, and code-switching, are not yet fully included. Future work can extend the taxonomy toward broader and more realistic instruction scenarios while preserving its diagnostic structure.

Second, the current multilingual setting is not fully balanced. English and Chinese use large splits, whereas the other languages are evaluated with mini splits under the same taxonomy. While this design enables broad multilingual coverage at manageable cost, future extensions should enlarge more languages into full-scale settings and improve cross-lingual comparability.

Third, the evaluation pipeline relies partly on LALM-based judging. Although its agreement with human judgments supports its scalability, such judges may still introduce bias, instability, and model-version dependence, while commercial APIs also raise concerns about cost and reproducibility. Future work should therefore strengthen robustness through broader human validation, multi-judge calibration, and more stable open-source evaluators.

Overall, we view MINT-Bench as a benchmark foundation rather than a final endpoint. We hope future efforts will expand its coverage, improve multilingual balance, and reduce dependence on proprietary evaluators, enabling more reproducible and diagnostically grounded progress in instruction-following TTS.

\newpage

\bibliographystyle{ACM-Reference-Format}
\bibliography{reference}

\newpage

\appendix

\twocolumn[{%
  \begin{center}
    \vspace{2em}
    \LARGE \bfseries APPENDICES
    \vspace{2em}
  \end{center}
}]

\section{Extended Details of the Hierarchical Multi-axis Taxonomy}
\label{sec:appendix_taxonomy}

\subsection{Expanded Definitions of Attribute Values}
\label{subsec:attribute_values}

As outlined in the main text, the core of MINT-Bench is grounded in ten atomic acoustic attributes that define the primary timbre and style inventory. To ensure structured control and reliable evaluation, we define an explicit, discrete set of core values for each attribute during the data construction phase. These precise tags serve as the semantic scaffolding for generating diverse natural language instructions. Table \ref{tab:attribute_values} provides the exhaustive list of all core values defined in MINT-Bench.

Note that the \textit{Accent} attribute is inherently language-specific. In the full evaluation split, MINT-Bench supports fine-grained accent control primarily for English and Chinese, capturing prominent regional and socio-linguistic variations.

\begin{table*}[t]
\caption{Exhaustive list of core values for the ten atomic timbre and style attributes used in MINT-Bench. These discrete tags are used during the Structured Label Planning stage to ensure controlled and balanced benchmark coverage.}
\label{tab:attribute_values}
\centering
\small
\renewcommand{\arraystretch}{1.3} 
\begin{tabularx}{\textwidth}{@{} >{\bfseries}l l X @{}}
\toprule
\textbf{Category} & \textbf{Attribute} & \textbf{Discrete Core Values} \\
\midrule

\multirow{4}{*}{\textbf{Timbre}} 
& Age & child, adult, elderly \\
\cmidrule{2-3}
& Gender & male, female \\
\cmidrule{2-3}
& Pitch & very\_high, high, normal, low, very\_low \\
\cmidrule{2-3}
& Texture & deep, rich, breathy, husky, smoky, magnetic, soft, bright, clear, nasal, sweet, mellow \\

\midrule

\multirow{8}{*}{\textbf{Style}} 
& Speed & very\_fast, fast, normal, slow, very\_slow \\
\cmidrule{2-3}
& Volume & very\_quiet, quiet, normal, loud, very\_loud \\
\cmidrule{2-3}
& Emotion & neutral, happy, angry, sad, fearful, surprised, worried, thoughtful \\
\cmidrule{2-3}
& Personality & cheerful, calm, serious, lively, reserved, humorous, gentle, confident, aloof, warm \\
\cmidrule{2-3}
& Tone & firm, euphemistic, hesitant, urging, sarcastic, complaining, coquettish, pleading, indifferent, playful, awkward, disappointed, speechless, disdainful, teasing, smug, defiant \\
\cmidrule{2-3}
& Accent (English) & american\_accent, british\_accent, indian\_accent \\
\cmidrule{2-3}
& Accent (Chinese) & putonghua, beijing\_accent, tianjin\_accent, sichuan\_accent, northeastern\_accent, taiwanese\_mandarin, cantonese\_accented\_mandarin, shanghai\_accented\_mandarin \\

\bottomrule
\end{tabularx}
\end{table*}

\subsection{Full Taxonomy Nodes, Control Patterns, and Budgets}
\label{subsec:taxonomy_nodes_budgets}

To provide a comprehensive view of MINT-Bench's internal structure, we detail the complete list of taxonomy nodes, fine-grained control patterns, representative examples, and exact instruction budgets for both the \textbf{Large} and \textbf{Mini} evaluation splits. Due to page constraints, this information is divided into two parts: Table \ref{tab:full_taxonomy_nodes_part1} covers the Easy and Hard levels, while Table \ref{tab:full_taxonomy_nodes_part2} presents the Special (extra-vocal) level alongside the overall dataset statistics.

The \textbf{Large} split is designed for exhaustive evaluation on primary languages (e.g., English and Chinese), comprising 890 instruction-text pairs in total. The \textbf{Mini} split serves as a lightweight, cost-effective alternative for broader multilingual evaluation, derived by sampling approximately 30\% of the full budget (with a minimum guarantee of 4 items per atomic leaf node to prevent under-representation), resulting in 274 pairs per language. 

For atomic timbre and style attributes, the budget is distributed proportionally across their respective sub-categories (e.g., Age, Gender, Pitch, Texture, Speed, Volume, Emotion, Accent, Personality, Tone). Similarly, the \textit{Character} node aggregates four distinct persona types: Social Roles, Stylized Personas, Non-IP Archetypes, and IP Portraits.

\begin{table*}[htbp]
\caption{The hierarchical taxonomy of MINT-Bench (Part 1/2), detailing the descriptions, prompt examples, and item budgets for the \textbf{Easy} and \textbf{Hard} levels. The examples illustrate how structured intents are transformed into natural language instructions.}
\label{tab:full_taxonomy_nodes_part1}
\centering
\small
\renewcommand{\arraystretch}{1.25}
\begin{tabularx}{\textwidth}{@{} >{\raggedright\arraybackslash}p{3.5cm} X c c @{}}
\toprule
\textbf{Taxonomy Node} & \textbf{Description \& Representative Example} & \textbf{Large} & \textbf{Mini} \\
\midrule

\multicolumn{4}{c}{\cellcolor{gray!10}\textbf{Easy Level}} \\
\midrule

\textbf{Easy / Timbre / Tag} \newline {\scriptsize (Age, Gender, Pitch, Texture)} 
& \textbf{Description:} Dictionary-like description of a single timbre tag. \newline
  \textbf{Example:} ``Age: child'' 
& 61 & 20 \\
\midrule

\textbf{Easy / Timbre / Direct} \newline {\scriptsize (Age, Gender, Pitch, Texture)} 
& \textbf{Description:} Straightforward natural language description of a single timbre tag. \newline
  \textbf{Example:} ``Speak in a child's voice.'' 
& 61 & 20 \\
\midrule

\textbf{Easy / Style / Tag} \newline {\scriptsize (Speed, Volume, Emotion, Accent, Personality, Tone)} 
& \textbf{Description:} Dictionary-like description of a single style tag. \newline
  \textbf{Example:} ``Emotion: Happy'' 
& 132 & 40 \\
\midrule

\textbf{Easy / Style / Direct} \newline {\scriptsize (Speed, Volume, Emotion, Accent, Personality, Tone)} 
& \textbf{Description:} Straightforward natural language description of a single style tag. \newline
  \textbf{Example:} ``Speak in a cheerful tone.'' 
& 132 & 40 \\
\midrule

\textbf{Easy / Composite / \newline Simple\_comp} 
& \textbf{Description:} Random pairs of timbre and/or style attributes in dictionary-like or straightforward format. \newline
  \textbf{Example:} ``Speed: very fast, Emotion: excited'' or ``A boy spoke in a sad tone.'' 
& 72 & 22 \\

\midrule
\multicolumn{4}{c}{\cellcolor{gray!10}\textbf{Hard Level}} \\
\midrule

\textbf{Hard / Composite / \newline Complex\_comp (Static)} 
& \textbf{Description:} Three or more attributes (timbre/style) combined in a single static prompt. \newline
  \textbf{Example:} ``Age: adult, Gender: female, Tone: joking, Volume: normal'' or ``A middle-aged woman joked in a normal voice volume.'' 
& 72 & 22 \\
\midrule

\textbf{Hard / Composite / \newline Dynamic} 
& \textbf{Description:} Dynamic attributes (pitch, speed, volume, emotion) that change or progress over time. \newline
  \textbf{Example:} ``Begin with uncertainty... slowly turn into tension... then fade into hopelessness...'' 
& 36 & 11 \\
\midrule

\textbf{Hard / Composite / \newline Layered} 
& \textbf{Description:} Multi-layered emotions or subtexts indicating psychological depth. \newline
  \textbf{Example:} ``Pretending to be indifferent, but actually caring deeply.'' 
& 32 & 10 \\
\midrule

\textbf{Hard / Composite / \newline Conflict} 
& \textbf{Description:} Conflicts and contrasts between different attributes (e.g., mismatched timbre and tone). \newline
  \textbf{Example:} ``Using a childlike, playful tone to deliver commands that radiate authority and pressure.'' 
& 32 & 10 \\
\midrule

\textbf{Hard / Persona / \newline Scenario} 
& \textbf{Description:} Prompts based on specific scenarios such as news broadcasts, poetry recitations, or bedtime stories. \newline
  \textbf{Example:} ``Gentle Mom Tells Bedtime Stories: Female voice, softer and lower pitch, slow pace...'' 
& 30 & 9 \\
\midrule

\textbf{Hard / Persona / \newline Character} \newline {\scriptsize (Social Role, Stylized Persona, Non-IP Archetype, IP Portrait)} 
& \textbf{Description:} Includes standard characters, stylized persona types, and vocal profiles of popular IPs. \newline
  \textbf{Example:} ``Please imitate a middle-aged man around 50 years old with a slightly overweight build, speaking in authentic Sichuan Mandarin...'' 
& 122 & 37 \\

\bottomrule
\end{tabularx}
\end{table*}

\begin{table*}[htbp]
\caption{The hierarchical taxonomy of MINT-Bench (Part 2/2), detailing the descriptions, prompt examples, and item budgets for the \textbf{Special} (extra-vocal) level, along with the total budget distributions.}
\label{tab:full_taxonomy_nodes_part2}
\centering
\small
\renewcommand{\arraystretch}{1.25}
\begin{tabularx}{\textwidth}{@{} >{\raggedright\arraybackslash}p{3.5cm} X c c @{}}
\toprule
\textbf{Taxonomy Node} & \textbf{Description \& Representative Example} & \textbf{Large} & \textbf{Mini} \\
\midrule

\multicolumn{4}{c}{\cellcolor{gray!10}\textbf{Special Level (Extra-Vocal)}} \\
\midrule

\textbf{Special / Abnormal / \newline Disfluency} 
& \textbf{Description:} Emphasizing awkward speech flow and lack of fluency. \newline
  \textbf{Example:} ``Slight stutter, with more noticeable hesitations when saying key words...'' 
& 20 & 6 \\
\midrule

\textbf{Special / Abnormal / \newline Dysphonia} 
& \textbf{Description:} Emphasizing abnormal vocalization states, such as breathing, vocal cords, or physical strain. \newline
  \textbf{Example:} ``Speaking as if just finishing a run, breathing unevenly...'' 
& 24 & 7 \\
\midrule

\textbf{Special / Nonverbal / \newline Implicit} 
& \textbf{Description:} Implicitly describing nonverbal events through text labels or brackets. \newline
  \textbf{Example:} \textit{Instruction:} ``The tone was extremely impatient...'', \textit{Text:} ``I can't take it anymore, [breath] I really can't.'' 
& 32 & 10 \\
\midrule

\textbf{Special / Nonverbal / \newline Explicit} 
& \textbf{Description:} Explicitly describing nonverbal behaviors directly within the instructions. \newline
  \textbf{Example:} ``The sound of breathing is quite noticeable, with a distinct inhale at the beginning of each sentence.'' 
& 32 & 10 \\

\midrule
\multicolumn{2}{r}{\textbf{Total Items per Language:}} & \textbf{890} & \textbf{274} \\
\bottomrule
\end{tabularx}
\end{table*}

\section{Details of the Data Construction Pipeline}
\label{sec:appendix_data_construction}

This section provides the exact prompts used to instruct the Large Language Models (LLMs) during the multi-stage data construction pipeline of MINT-Bench. To ensure reproducibility and accommodate the double-column layout, the complete prompts for both the Structured Label Planning stage and the Instruction-Text Pair Construction stage are presented in Figure \ref{fig:prompt_label_planning_part1}, Figure \ref{fig:prompt_label_planning_part2}, Figure \ref{fig:prompt_instruction_text_part1} and Figure \ref{fig:prompt_instruction_text_part2}, respectively.

\subsection{Prompts for Structured Label Planning}
\label{subsec:prompt_label_planning}

In the first stage, the LLM acts as a planning assistant to generate a clean, diverse, and non-leaking structured label plan for a given taxonomy node. The prompt enforces strict constraints on text semantics, length, language policies, and taxonomy-specific formatting. Due to its comprehensive nature, the full prompt is presented in two parts: Figure \ref{fig:prompt_label_planning_part1} outlines the general generation rules, multi-lingual policies, and the expected JSON output structure, while Figure \ref{fig:prompt_label_planning_part2} details the fine-grained field definitions, text length constraints, and node-specific realization rules.

\begin{figure*}[htbp]
\centering
\begin{tcolorbox}[
    width=\textwidth,
    colback=gray!2, 
    colframe=black!70, 
    title=Prompt: Structured Label Planning (Part 1/2), 
    fonttitle=\bfseries,
    boxrule=0.8pt,
    arc=3pt
]
\footnotesize
You are helping construct a benchmark for instruction-following text-to-speech (TTS). \\
Your task in this stage is NOT to generate final instruction-text pairs. \\
Instead, generate a structured label plan for a given taxonomy node.

Each plan item should specify: \\
1. the taxonomy path, \quad 2. the control target(s), \quad 3. the intended target value(s), \\
4. the realization style needed later, \quad 5. the intended text semantic type and text length, \\
6. any constraints needed later for generating the final instruction and text.

The goal of this stage is to create a clean, diverse, balanced, non-leaking, benchmark-ready scratch plan that can later be converted into final instruction-text pairs.

\vspace{0.5em}
\textbf{--- GENERAL RULES ---} \\
1. Do NOT generate final polished instructions. Only plan the later instruction style and constraints. \\
2. Do NOT generate final text sentences. Only specify text\_semantic, text\_length, and text\_constraints. \\
3. Text semantics must not trivially leak the target control attributes. \\
\hspace*{1em} Examples: \\
\hspace*{1em} - if the target is age=child, do not force child-themed lexical content \\
\hspace*{1em} - if the target is emotion=angry, do not force obviously angry words into the text \\
\hspace*{1em} - if the target is accent=sichuan, do not force dialect words unless explicitly required \\
4. Keep the difficulty determined mainly by the taxonomy node, not by overly complicated text content. \\
5. Prefer diversity across: target values, semantic content types, attribute combinations, and instruction realization styles when the node allows more than one realization style. \\
6. Use the provided allowed\_labels and allowed\_values strictly. Do not invent out-of-scope labels or values. \\
7. Output ONLY valid JSON. No explanations outside JSON.

\vspace{0.5em}
\textbf{--- LANGUAGE REALIZATION POLICY ---} \\
A. Primary languages \\
- If language is zh, the later final instruction and final text must both be Chinese. \\
- If language is en, the later final instruction and final text must both be English. \\
B. Non-primary languages \\
- If language is ja / ko / pt / fr / de / it / ru / es: \\
\hspace*{1em} - the later final instruction must be written in English \\
\hspace*{1em} - the later final instruction should explicitly cue the target language with a phrase like: "Speak in Japanese." \\
\hspace*{1em} - the later final text must be written in the target language itself \\
- Plan items must respect this downstream realization rule.

\vspace{0.5em}
\textbf{--- OUTPUT FORMAT ---} \\
Return a JSON object with the following structure: \\
\{ \\
\hspace*{1em} "taxonomy\_node": "$<$input taxonomy node$>$", \\
\hspace*{1em} "language": "$<$input language$>$", \\
\hspace*{1em} "generation\_summary": \{ \\
\hspace*{2em} "num\_items": $<$int$>$, "domain": "$<$timbre|style|composite|persona|extra\_vocal$>$", \\
\hspace*{2em} "difficulty": "$<$easy|hard|special$>$", "notes": "$<$short note$>$" \\
\hspace*{1em} \}, \\
\hspace*{1em} "items": [ \\
\hspace*{2em} \{ \\
\hspace*{3em} "plan\_id": "$<$unique short id$>$", "meta\_path": "$<$taxonomy path$>$", \\
\hspace*{3em} "attribute\_set": ["$<$attr1$>$", "$<$attr2$>$"], "instruction\_form": "$<$tag|direct$>$", \\
\hspace*{3em} "target\_values": \{ "$<$attr1$>$": "$<$value1$>$", "$<$attr2$>$": "$<$value2$>$" \}, \\
\hspace*{3em} "aux\_values": \{ "$<$aux\_key$>$": "$<$aux\_value$>$" \}, \\
\hspace*{3em} "text\_semantic": "$<$semantic type$>$", "text\_length": "$<$short|medium|long$>$", \\
\hspace*{3em} "text\_constraints": [ "$<$constraint 1$>$" ], "instruction\_constraints": [ "$<$constraint 1$>$" ], \\
\hspace*{3em} "flags": \{ \\
\hspace*{4em} "static\_or\_dynamic": "$<$static|dynamic$>$", "requires\_roleplay": $<$true|false$>$, \\
\hspace*{4em} "requires\_nonverbal\_tag": $<$true|false$>$, "requires\_abnormal\_state": $<$true|false$>$, \\
\hspace*{4em} "contains\_conflict": $<$true|false$>$, "contains\_layered\_affect": $<$true|false$>$ \\
\hspace*{3em} \} \\
\hspace*{2em} \} \\
\hspace*{1em} ] \\
\}
\end{tcolorbox}
\caption{The prompt template for Structured Label Planning (Part 1). This part establishes the general generation boundaries, multi-lingual control policies, and the strict JSON output structure.}
\label{fig:prompt_label_planning_part1}
\end{figure*}

\begin{figure*}[htbp]
\centering
\begin{tcolorbox}[
    width=\textwidth,
    colback=gray!2, 
    colframe=black!70, 
    title=Prompt: Structured Label Planning (Part 2/2), 
    fonttitle=\bfseries,
    boxrule=0.8pt,
    arc=3pt
]
\footnotesize
\textbf{--- FIELD DEFINITIONS ---} \\
- instruction\_form: \textbf{tag} (fixed dictionary-like tags) or \textbf{direct} (natural voice-direction phrasing). \\
- tag format rule: Chinese uses "shuxing: zhi"; non-Chinese uses "Attribute: Value". For multi-tag items, each attribute must follow the same fixed pattern.

\vspace{0.5em}
\textbf{--- TEXT LENGTH PLANNING RULES ---} \\
- text\_length: \textbf{short} (brief but complete utterance), \textbf{medium} (one/two full sentences), \textbf{long} (staged delivery/dynamic progression). \\
- Atomic easy nodes may use short or medium. \\
- easy/composite/simple\_comp/static should be mostly medium (avoid overly tiny text for two-attribute control). \\
- hard/composite/complex\_comp/static should be medium or long. \\
- dynamic nodes should be long. \\
- layered, conflict, persona, abnormal, implicit\_nv, and explicit\_nv should usually be medium or long.

\vspace{0.5em}
\textbf{--- NODE-SPECIFIC RULES ---} \\
A. \textbf{easy / timbre \& style}: exactly one attribute, instruction\_form must match tag/direct. \\
B. \textbf{easy / composite / simple\_comp / static}: exactly two attributes. No conflict, layered affect, or dynamic progression. Balanced mix of instruction forms (tag \& direct). \\
C. \textbf{hard / composite / complex\_comp / static}: three or more attributes. Mostly direct items, meaningful minority of tag items. \\
D. \textbf{hard / composite / complex\_comp / dynamic}: one or more attributes must change over time. instruction\_form direct. Prefer multi-stage progression. \\
E. \textbf{hard / composite / complex\_comp / layered}: multi-layered emotion or hidden affect. instruction\_form direct. Text allows dual interpretation. \\
F. \textbf{hard / composite / complex\_comp / conflict}: conflicting or counter-stereotypical cues. instruction\_form direct. Text remains relatively simple. \\
G. \textbf{persona nodes}: instruction\_form direct. For ip\_portrait, aux\_values store internal anchors (franchise, character\_name) that must NOT be copied directly into the final user-facing instruction. \\
H. \textbf{special / extra\_vocal / abnormal}: instruction\_form direct. Treat as continuous speaking states, no explicit nonverbal tags in future text. \\
I. \textbf{special / extra\_vocal / implicit\_nv}: instruction\_form direct. Future instruction describes nonverbal events; future text MUST NOT contain explicit bracket tags. \\
J. \textbf{special / extra\_vocal / explicit\_nv}: instruction\_form direct. Future text MUST contain explicit bracket tags. Future instruction should NOT mention them, but only control global delivery style.

\vspace{0.5em}
\textbf{--- TEXT SEMANTIC EXAMPLES ---} \\
Possible text\_semantic values include: daily\_notice, casual\_statement, simple\_request, factual\_statement, short\_reflection, mild\_dialogue, imperative\_command, emotionally\_ambiguous\_statement, narratable\_sentence, etc.

\vspace{0.5em}
\textbf{--- INPUT VARIABLES ---} \\
You will be given: taxonomy\_node, num\_items, allowed\_labels, allowed\_values, optional aux\_value\_lookup, language, language\_profile. \\
Use them strictly. Now generate the label plan.
\end{tcolorbox}
\caption{The prompt template for Structured Label Planning (Part 2). This part defines node-specific parsing logic, fine-grained text length constraints, and extra paralinguistic configurations.}
\label{fig:prompt_label_planning_part2}
\end{figure*}

\subsection{Prompts for Instruction-Text Pair Construction}
\label{subsec:prompt_instruction_text}

In the second stage, the model converts the structured label plans into natural, usable instruction-text pairs. The prompt explicitly specifies language requirements, handles bracketed non-verbal cues for special nodes, and generates multiple candidates per plan to ensure diversity. Due to its comprehensive nature, the full prompt is presented in two parts: Figure \ref{fig:prompt_instruction_text_part1} details the core generation rules, language policies, and strict formatting for tag and direct instructions, while Figure \ref{fig:prompt_instruction_text_part2} outlines the node-specific constraints and the expected JSON output structure.

\begin{figure*}[htbp]
\centering
\begin{tcolorbox}[
    width=\textwidth,
    colback=gray!2, 
    colframe=black!70, 
    title=Prompt: Instruction-Text Pair Construction (Part 1/2), 
    fonttitle=\bfseries,
    boxrule=0.8pt,
    arc=3pt
]
\footnotesize
You are helping construct a benchmark for instruction-following text-to-speech (TTS). \\
Your task is to convert a structured label plan into final instruction-text pairs.

For each plan item, generate EXACTLY \{num\_candidates\} candidate instruction-text pairs. \\
The candidates do not need to be maximally different from one another. They may vary naturally in wording, phrasing, or text content. The goal is to produce faithful, usable candidates and then let the pipeline randomly select one.

\vspace{0.5em}
\textbf{--- GENERAL RULES ---} \\
1. Follow the label plan strictly. Do not change meta\_path. Do not add extra control targets. Do not ignore instruction\_form, text\_semantic, or instruction\_constraints. \\
2. The instruction must match the intended realization mechanism: \textbf{tag} (fixed key-value tags) or \textbf{direct} (natural voice-direction). \\
3. The text must be natural and speakable for TTS. Avoid fragments unless explicitly required. Avoid obscure wording. \\
4. The text must NOT trivially leak the target. \\
\hspace*{1em} Examples: \\
\hspace*{1em} - child voice target does not mean child-themed lexical content \\
\hspace*{1em} - angry target does not mean explicit angry wording \\
\hspace*{1em} - accent target does not mean dialect spellings unless explicitly required \\
5. Difficulty should come mainly from the control design, not from making the text content artificially complicated. \\
6. Respect aux\_values only as interpretation hints. Do NOT copy franchise names, titles, or character names into the final instruction or text. \\
7. Output ONLY valid JSON.

\vspace{0.5em}
\textbf{--- LANGUAGE POLICY ---} \\
A. zh: instruction must be Chinese, text must be Chinese. \\
B. en: instruction must be English, text must be English. \\
C. ja / ko / pt / fr / de / it / ru / es: \\
\hspace*{1em} - instruction must be English \\
\hspace*{1em} - instruction must explicitly cue the target language with a prefix like: "Speak in Japanese." \\
\hspace*{1em} - text must be written in the target language itself

\vspace{0.5em}
\textbf{--- TAG INSTRUCTION RULES ---} \\
If instruction\_form is tag, follow these strictly: \\
1. Chinese tag instructions must use fixed "shuxing: zhi" format. \\
2. All non-Chinese instruction languages must use fixed "Attribute: Value" format. \\
3. For multi-attribute tag items, every attribute must appear in the same fixed format. \\
4. Do not paraphrase tag items into free prose. \\
5. For non-primary languages, keep the English target-language prefix and then use English key-value tags. \\
\hspace*{1em} Example pattern: "Speak in Japanese. Attribute: Value, Attribute: Value"

\vspace{0.5em}
\textbf{--- DIRECT INSTRUCTION RULES ---} \\
If instruction\_form is direct: \\
- use natural, operational voice-direction phrasing \\
- keep easy items concise, make hard items richer only when the control complexity truly needs it \\
- for ip\_portrait, use a richer and more distinctive direction style than ordinary persona items
\end{tcolorbox}
\caption{The prompt template for Instruction-Text Pair Construction (Part 1). This part establishes the overarching generation rules, multi-lingual handling, and strict stylistic boundaries for tag-based and direct instructions.}
\label{fig:prompt_instruction_text_part1}
\end{figure*}

\begin{figure*}[htbp]
\centering
\begin{tcolorbox}[
    width=\textwidth,
    colback=gray!2, 
    colframe=black!70, 
    title=Prompt: Instruction-Text Pair Construction (Part 2/2), 
    fonttitle=\bfseries,
    boxrule=0.8pt,
    arc=3pt
]
\footnotesize
\textbf{--- NODE-SPECIFIC RULES ---} \\
A. easy / composite / simple\_comp / static: exactly two attributes. Obey instruction\_form. \\
B. hard / composite / complex\_comp / static: three or more attributes. For direct, integrate coherently; for tag, keep strict key-value format. \\
C. hard / composite / complex\_comp / dynamic: describe temporal change clearly and in order. \\
D. hard / composite / complex\_comp / layered: instruction makes layered reading clear; text supports dual interpretation. \\
E. hard / composite / complex\_comp / conflict: instruction makes contradiction vivid; text stays relatively plain. \\
F. persona: direct-form instructions. ip\_portrait must not copy copyrighted names from aux\_values, convert them into distinctive abstract vocal portraits. \\
G. abnormal: disfluency and dysphonia are continuous performance directions, not bracket tags. \\
H. implicit\_nv: instruction describes nonverbal vocal events, but text MUST NOT contain explicit bracket tags. \\
I. explicit\_nv: text MUST contain explicit bracketed nonverbal tags. Instruction must NOT mention specific nonverbal events or bracket tags (describe global speaking manner/emotion instead).

\vspace{0.5em}
\textbf{--- OUTPUT FORMAT ---} \\
Return a JSON object with the following structure: \\
\{ \\
\hspace*{1em} "results": [ \\
\hspace*{2em} \{ \\
\hspace*{3em} "plan\_id": "$<$copied from input$>$", \\
\hspace*{3em} "meta\_path": "$<$copied from input$>$", \\
\hspace*{3em} "target\_values": \{ "$<$attr1$>$": "$<$value1$>$" \}, \\
\hspace*{3em} "candidates": [ \\
\hspace*{4em} \{ \\
\hspace*{5em} "candidate\_id": "c1", \\
\hspace*{5em} "instruction": "$<$candidate instruction$>$", \\
\hspace*{5em} "text": "$<$candidate text$>$", \\
\hspace*{5em} "quality\_checks": \{ \\
\hspace*{6em} "matches\_plan": true, \\
\hspace*{6em} "text\_leaks\_target": false, \\
\hspace*{6em} "contains\_explicit\_nonverbal\_tag": $<$true|false$>$, \\
\hspace*{6em} "notes": "$<$short note$>$" \\
\hspace*{5em} \} \\
\hspace*{4em} \} \\
\hspace*{3em} ] \\
\hspace*{2em} \} \\
\hspace*{1em} ] \\
\} \\
Return JSON only.
\end{tcolorbox}
\caption{The prompt template for Instruction-Text Pair Construction (Part 2). This part details the specialized realization constraints for complex composition and paralinguistic extra-vocal nodes, along with the expected JSON schema.}
\label{fig:prompt_instruction_text_part2}
\end{figure*}

\section{Extended Details of the Evaluation Protocol}
\label{sec:appendix_evaluation}

\subsection{LALM Prompts for Evaluation Pipeline}
\label{subsec:prompt_evaluation}

In the hierarchical hybrid evaluation protocol, MINT-Bench employs a Large Audio Language Model (LALM) as an automatic judge to assess instruction-following and perceptual quality. Crucially, we decouple content consistency (evaluated via objective WER thresholds) from vocal controllability. Therefore, the LALM is explicitly instructed to ignore minor lexical mismatches and focus entirely on whether the acoustic characteristics align with the prompt.

Because the diagnostic criteria vary significantly across different structural complexities, we design three tailored evaluation prompts corresponding to the \textbf{Easy}, \textbf{Hard}, and \textbf{Special} subsets. Each prompt guides the LALM to assign a 3-level instruction-following score alongside conditional boolean bonuses for naturalness and expressiveness. The complete evaluation prompts for Easy, Hard, and Special nodes are presented in Figure \ref{fig:prompt_eval_easy}, Figure \ref{fig:prompt_eval_hard}, and Figure \ref{fig:prompt_eval_special}, respectively.

\begin{figure*}[htbp]
\centering
\begin{tcolorbox}[
    width=\textwidth,
    colback=blue!2, 
    colframe=black!70, 
    title=Evaluation Prompt: Easy Nodes, 
    fonttitle=\bfseries,
    boxrule=0.8pt,
    arc=3pt
]
\footnotesize
You are an evaluator for an instruction-following text-to-speech benchmark. \\
Your task is to evaluate ONLY whether the audio follows the requested speaking instruction. \\
The audio has already passed WER-based content screening. Therefore: \\
- Do NOT evaluate transcript accuracy. \\
- Do NOT penalize minor wording mismatch. \\
- Do NOT score based on lexical fidelity. \\
- Focus only on audible instruction-following.

You will be given: taxonomy\_node, plan\_id, language, instruction, text, target\_values, and one audio sample. \\
Use the natural-language instruction as the primary evaluation target. Use target\_values only as a structured anchor to identify which dimensions must be checked. If the instruction is more specific than target\_values, always evaluate against the instruction.

\vspace{0.5em}
\textbf{--- TASK SCOPE ---} \\
This prompt is for EASY nodes only (e.g., easy/timbre/tag/atomic/*, easy/composite/simple\_comp/static). \\
For atomic nodes: evaluate exactly one requested attribute. \\
For easy composite static nodes: evaluate exactly the requested attributes in target\_values. Do not add extra criteria.

\vspace{0.5em}
\textbf{--- EVALUATION ATTITUDE ---} \\
Be conservative and evidence-based. Do NOT assume a style is present just because it would be plausible. If evidence is weak, ambiguous, unstable, or too mild, do not give full credit. \\
Important: \\
- gender and age must be judged very carefully \\
- accent must not be credited unless clearly audible \\
- strong labels such as low/high/loud/elderly/breathy must be clearly audible, not merely approximate \\
- if a key requested attribute clearly conflicts with the audio, instruction-following cannot receive level 3

\vspace{0.5em}
\textbf{--- INSTRUCTION-FOLLOWING LEVELS ---} \\
Assign one instruction-following level: \\
\textbf{1 = bad}: the requested attribute(s) are absent, clearly contradicted, or mostly failed \\
\textbf{2 = so-so}: some target attribute(s) are present, but weak, incomplete, insufficient in degree, or partially contradicted \\
\textbf{3 = good}: all requested attribute(s) are clearly and convincingly realized; no major contradiction; degree/intensity is sufficient

\vspace{0.5em}
\textbf{--- BONUS RULES ---} \\
Only if instruction-following level = 3, evaluate bonus items: \\
\textbf{Naturalness bonus}: true if the speech sounds humanlike, smooth, and free of obvious synthetic stiffness/artifacts. \\
\textbf{Expressiveness bonus}: true if the delivery has clear, appropriate, and non-flat expressive variation. \\
If instruction-following level is 1 or 2, both bonuses must be false. \\
Final score = instruction\_following\_level + naturalness\_bonus + expressiveness\_bonus (Resulting in 1, 2, 3, 4, or 5).

\vspace{0.5em}
\textbf{--- OUTPUT ---} \\
Return ONLY valid JSON in this format: \\
\{ \\
\hspace*{1em} "plan\_id": "$<$plan\_id$>$", "taxonomy\_node": "$<$taxonomy\_node$>$", \\
\hspace*{1em} "instruction\_following\_level": 1, "instruction\_following\_label": "bad", \\
\hspace*{1em} "naturalness\_bonus": false, "expressiveness\_bonus": false, "final\_score": 1, \\
\hspace*{1em} "brief\_reason": "$<$short overall judgment$>$", \\
\hspace*{1em} "instruction\_evidence": \{ "$<$dim1$>$": "$<$brief evidence$>$" \}, \\
\hspace*{1em} "bonus\_evidence": \{ "naturalness": "$<$short note$>$", "expressiveness": "$<$short note$>$" \} \\
\} \\
Allowed instruction\_following\_label values: "bad", "so-so", "good". Return JSON only. \\
INPUT: $<$INPUT\_EVAL\_JSON$>$
\end{tcolorbox}
\caption{The LALM evaluation prompt used for Easy taxonomy nodes, focusing on conservative verification of atomic and simple composite attributes.}
\label{fig:prompt_eval_easy}
\end{figure*}

\begin{figure*}[htbp]
\centering
\begin{tcolorbox}[
    width=\textwidth,
    colback=red!2, 
    colframe=black!70, 
    title=Evaluation Prompt: Hard Nodes, 
    fonttitle=\bfseries,
    boxrule=0.8pt,
    arc=3pt
]
\footnotesize
You are an evaluator for an instruction-following text-to-speech benchmark. \\
Your task is to evaluate ONLY whether the audio follows the requested speaking instruction. \\
The audio has already passed WER-based content screening. (Do NOT evaluate transcript accuracy, lexical fidelity, etc.)

\vspace{0.5em}
\textbf{--- TASK SCOPE ---} \\
This prompt is for HARD nodes only (e.g., hard/composite/complex\_comp/*, hard/composite/persona/*).

\vspace{0.5em}
\textbf{--- GENERAL HARD-NODE PRINCIPLES ---} \\
For hard nodes, do not judge only by whether one local cue is present. Judge whether the intended multi-dimensional or structured target is truly realized. Be conservative: weak hints are not enough; flattened execution is not enough; a single successful attribute does not justify a good rating if the structure fails.

\vspace{0.5em}
\textbf{--- NODE-SPECIFIC RULES ---} \\
\textbf{1. complex\_comp/static}: judge whether multiple requested attributes are simultaneously realized. If only partial, assign level 2. \\
\textbf{2. complex\_comp/dynamic}: judge temporal progression explicitly. Check if multiple stages are audible, order matches, and progression is clear. If too weak or binary, do not give level 3. \\
\textbf{3. complex\_comp/layered}: judge whether both surface and hidden affect are audible. \\
\textbf{4. complex\_comp/conflict}: judge whether both conflicting cues are present. \\
\textbf{5. persona (scenario/social\_role/archetype)}: judge the overall role/scenario impression. Do not reward generic style if the requested persona is not clearly formed. \\
\textbf{6. persona/ip\_portrait}: evaluate against the abstract vocal portrait described in the instruction. Full credit requires the audio strongly evokes the described portrait through audible cues (age, texture, pitch, rhythm, accent, temperament). Vague generic resemblance is not enough for level 3.

\vspace{0.5em}
\textbf{--- INSTRUCTION-FOLLOWING LEVELS ---} \\
\textbf{1 = bad}: the requested structure, role, progression, or portrait is largely absent, contradicted, or poorly realized. \\
\textbf{2 = so-so}: some elements are present, but incomplete, flattened, weak, or structurally imperfect. \\
\textbf{3 = good}: the requested hard target is clearly realized. For dynamic/layered/conflict, internal structure is clear. For persona, impression is specific, recognizable, and convincing.

\vspace{0.5em}
\textbf{--- BONUS RULES \& OUTPUT ---} \\
(Bonus rules and JSON output schema follow the exact same format as the Easy prompt, returning a 3-level score, boolean bonuses, and structured evidence dicts.) \\
INPUT: $<$INPUT\_EVAL\_JSON$>$
\end{tcolorbox}
\caption{The LALM evaluation prompt used for Hard taxonomy nodes. It introduces strict guidelines for structural realization, such as evaluating temporal progressions, layered subtexts, and composite persona representations.}
\label{fig:prompt_eval_hard}
\end{figure*}

\begin{figure*}[htbp]
\centering
\begin{tcolorbox}[
    width=\textwidth,
    colback=green!2, 
    colframe=black!70, 
    title=Evaluation Prompt: Special Nodes (Extra-Vocal), 
    fonttitle=\bfseries,
    boxrule=0.8pt,
    arc=3pt
]
\footnotesize
You are an evaluator for an instruction-following text-to-speech benchmark. \\
Your task is to evaluate ONLY whether the audio follows the requested speaking instruction. \\
The audio has already passed WER-based content screening. (Do NOT evaluate transcript accuracy, lexical fidelity, etc.)

\vspace{0.5em}
\textbf{--- TASK SCOPE ---} \\
This prompt is for SPECIAL nodes only (special/extra\_vocal/abnormal/* and special/extra\_vocal/*\_nv).

\vspace{0.5em}
\textbf{--- GENERAL SPECIAL-NODE PRINCIPLES ---} \\
Be highly conservative. Do NOT mistake the following for valid special events or abnormal states: background noise, friction noise, clipping, reverb artifacts, random breath noise from recording, accidental mouth noise, one-off roughness, isolated hesitation, or cut/edit glitches. A special effect counts only if it is clearly produced by the speaker and is strong enough to be intentionally perceived.

\vspace{0.5em}
\textbf{--- NODE-SPECIFIC RULES ---} \\
\textbf{1. explicit\_nv / implicit\_nv} \\
Evaluate in two steps: \\
Step A: event existence - first judge whether the requested nonverbal event actually occurs. If it does not clearly occur, it cannot receive high credit. \\
Step B: placement / frequency / appropriateness - if the event occurs, judge whether its placement/frequency roughly matches the instruction, and its realization is clear and intentional. \\
Scoring implications: no clear event -> level 1; event exists but placement/intensity imperfect -> level 2; event clearly exists and well placed -> level 3.

\textbf{2. abnormal/disfluency} \\
Only count clearly speaker-produced fluency disruption. The disfluency must be sustained enough or patterned enough to be an intentional abnormal speaking state. Do not count a single irregularity or isolated hesitation as full realization.

\textbf{3. abnormal/dysphonia} \\
Only count a continuous abnormal voice-state condition. The state must be audibly sustained, not a single rough syllable or isolated artifact. Do not confuse ordinary texture with pathological/abnormal phonation unless clearly sustained.

\vspace{0.5em}
\textbf{--- INSTRUCTION-FOLLOWING LEVELS ---} \\
\textbf{1 = bad}: the requested event/state is absent, unclear, accidental, contradicted, or mostly not realized. \\
\textbf{2 = so-so}: the requested event/state is present to some extent, but weak, unstable, insufficiently clear, insufficiently continuous, or mismatched in position/frequency/pattern. \\
\textbf{3 = good}: the requested event/state is clearly and intentionally realized. For nonverbal nodes: event clearly occurs and is reasonably matched. For abnormal nodes: abnormal state is clearly sustained and instruction-consistent.

\vspace{0.5em}
\textbf{--- BONUS RULES \& OUTPUT ---} \\
(Bonus rules and JSON output schema follow the exact same format as the Easy prompt, returning a 3-level score, boolean bonuses, and structured evidence dicts.) \\
INPUT: $<$INPUT\_EVAL\_JSON$>$
\end{tcolorbox}
\caption{The LALM evaluation prompt used for Special (Extra-Vocal) taxonomy nodes. It applies a two-step verification process to rigorously distinguish intentional nonverbal events and sustained abnormal vocalizations from audio artifacts.}
\label{fig:prompt_eval_special}
\end{figure*}

\subsection{Language-Specific WER Thresholds}
\label{subsec:wer_thresholds}

As described in the hierarchical hybrid evaluation protocol (Section \ref{sec:evaluation_methodology}), MINT-Bench applies a language-specific Word Error Rate (WER) threshold to assess content consistency. Because Automatic Speech Recognition (ASR) performance varies significantly across different languages due to model capabilities and linguistic characteristics, applying a uniform threshold across all languages would disproportionately penalize low-resource languages. 

To ensure a fair evaluation, we compute a trimmed mean WER for each language by collecting the overall WER values of all evaluated systems, removing the highest and lowest extremes, and averaging the remaining scores. A synthesized sample is considered to have preserved the intended linguistic content if its sample-level WER falls below this language-specific threshold.

Table \ref{tab:wer_thresholds} presents the exact trimmed mean WER thresholds applied in our multilingual evaluation.

\begin{table*}[htbp]
\caption{Language-specific Word Error Rate (WER) thresholds used for content consistency checking. These thresholds are calculated as the trimmed mean of the overall WER scores across all evaluated systems per language.}
\label{tab:wer_thresholds}
\centering
\small
\renewcommand{\arraystretch}{1.2}
\begin{tabular}{lc}
\toprule
\textbf{Language} & \textbf{Trimmed Mean WER Threshold} \\
\midrule
Chinese (zh) & 2.35\% \\
English (en) & 3.63\% \\
Spanish (es) & 10.53\% \\
Italian (it) & 12.45\% \\
Portuguese (pt) & 16.86\% \\
German (de) & 17.57\% \\
French (fr) & 19.27\% \\
Japanese (ja) & 35.83\% \\
Russian (ru) & 36.14\% \\
Korean (ko) & 46.71\% \\
\bottomrule
\end{tabular}
\end{table*}

\subsection{Human Annotation Setup}
\label{subsec:human_annotation}

To validate the reliability of our LALM-based automatic judging pipeline and assess the true perceptual quality of the synthesized speech, we conducted a comprehensive human annotation study. To facilitate this, we developed a customized, web-based blind-testing platform tailored specifically for instruction-following TTS evaluation. 

Figure \ref{fig:human_annot_platform} presents a representative screenshot of the annotation interface. During the evaluation, annotators are presented with the natural language instruction and the reference synthesis text. They can then listen to side-by-side audio samples generated by different models (anonymized as ``model 1'', ``model 2'', etc.). For each model, annotators are required to provide a 3-level rating for \textit{Instruction Following} (Bad, So-so, Good), and optionally award binary bonus points for \textit{Expressiveness} and \textit{Naturalness}, strictly adhering to the same criteria provided to the LALM judge. The interface is designed to be intuitive and structured, ensuring that human evaluators can efficiently and consistently apply the hierarchical evaluation guidelines.

\begin{figure*}[htbp]
\centering
\includegraphics[width=\textwidth]{figs/human_annot.png}
\caption{Screenshot of the custom web-based human annotation platform. The interface displays the instruction, reference text, and side-by-side audio playback controls, allowing annotators to blindly rate models on instruction following, expressiveness, and naturalness.}
\label{fig:human_annot_platform}
\end{figure*}

\section{Additional Experimental Results}
\label{sec:appendix_experiments}

\subsection{Full Multilingual Evaluation Results}
\label{subsec:full_multilingual_results}

In the main text (Section ~\ref{sec:experiment}), Table ~\ref{tab:main_results_en_zh_full} presented the detailed, node-level evaluation results exclusively for the primary languages (English and Chinese) due to page limits, while Table ~\ref{tab:multilingual_leaderboard_full} provided only an aggregated view of the Perceptual Quality (PQ) scores across all ten languages. 

To provide a complete and transparent view of model performance across diverse linguistic settings, this section presents the exhaustive evaluation results for the remaining eight languages: Japanese, Korean, French, German, Spanish, Portuguese, Italian, and Russian. 

Tables \ref{tab:main_results_ja} through \ref{tab:main_results_ru} detail the objective Word Error Rate (WER), the overarching Instruction-Following (IF) scores, and the Perceptual Quality (PQ) scores broken down by the Easy, Hard, and Special structural complexity levels. Consistent with the primary results, models are categorized into commercial and open-source systems. Please note that the Timbre Diversity Score (TDS) relies on a larger sample pool; thus, it is omitted from these extended mini-split tables to prevent skewed metric interpretations.

\begin{table*}[t]
\centering
\small
\setlength{\tabcolsep}{3.8pt}
\renewcommand{\arraystretch}{1.1}

\providecommand{\scorecell}[1]{#1}
\providecommand{\bestscore}[1]{\textbf{#1}}
\providecommand{\secondscore}[1]{\underline{#1}}
\providecommand{\pairscore}[2]{\makecell[c]{#1 / #2}}

\caption{Detailed results on Japanese (ja). Scores for Overall, Timbre, and Style are reported as \textit{Instruction-Following (IF) score / Perceptual Quality (PQ) score}. Fine-grained columns report the PQ score.}
\label{tab:main_results_ja}

\resizebox{\textwidth}{!}{
\begin{tabular}{>{\raggedright\arraybackslash\bfseries\normalsize}l l *{17}{c}}
\toprule
\multirow{3}{*}{\normalsize\bfseries Model} 
& \multirow{3}{*}{\normalsize\bfseries Lang} 
& \multirow{3}{*}{\normalsize\bfseries WER $\downarrow$} 
& \multirow{3}{*}{\normalsize\bfseries Overall $\uparrow$} 
& \multirow{3}{*}{\normalsize\bfseries Timbre $\uparrow$} 
& \multirow{3}{*}{\normalsize\bfseries Style $\uparrow$} 
& \multicolumn{3}{c}{\normalsize\bfseries Easy $\uparrow$} 
& \multicolumn{6}{c}{\normalsize\bfseries Hard $\uparrow$} 
& \multicolumn{4}{c}{\normalsize\bfseries Special $\uparrow$} \\
\cmidrule(lr){7-9} \cmidrule(lr){10-15} \cmidrule(lr){16-19}
& & & & & 
& \footnotesize\bfseries Tag & \footnotesize\bfseries Direct & \footnotesize\bfseries B-Comp.
& \footnotesize\bfseries M-Comp. & \footnotesize\bfseries Dynamic & \footnotesize\bfseries Layered & \footnotesize\bfseries Conflict & \footnotesize\bfseries Scenario & \footnotesize\bfseries Character
& \footnotesize\bfseries Disfl. & \footnotesize\bfseries Dysph. & \footnotesize\bfseries Impl. & \footnotesize\bfseries Expl. \\
\midrule

\rowcolor{tableblue}
\multicolumn{19}{c}{\normalsize\bfseries Commercial Systems} \\
\midrule
Gemini 2.5-Pro & ja & 2.9\% & \pairscore{2.67}{4.05} & \pairscore{2.63}{3.86} & \pairscore{2.85}{4.39} & \scorecell{4.08} & \scorecell{3.98} & \scorecell{4.41} & \scorecell{4.36} & \scorecell{4.64} & \scorecell{4.20} & \scorecell{4.56} & \scorecell{4.22} & \scorecell{3.60} & \scorecell{2.67} & \scorecell{3.86} & \scorecell{3.90} & \scorecell{4.20} \\
Gemini 2.5-Flash & ja & 3.1\% & \pairscore{2.67}{4.03} & \pairscore{2.79}{4.11} & \pairscore{2.82}{4.35} & \scorecell{4.30} & \scorecell{4.17} & \scorecell{4.32} & \scorecell{4.41} & \scorecell{4.00} & \scorecell{3.90} & \scorecell{4.30} & \scorecell{3.89} & \scorecell{3.55} & \scorecell{2.00} & \scorecell{3.86} & \scorecell{4.30} & \scorecell{3.40} \\
ElevenLabs-ttv-v3 & ja & 11.0\% & \pairscore{2.44}{3.39} & \pairscore{2.77}{3.52} & \pairscore{2.47}{3.50} & \scorecell{3.48} & \scorecell{3.43} & \scorecell{3.89} & \scorecell{3.62} & \scorecell{3.08} & \scorecell{3.19} & \scorecell{2.64} & \scorecell{3.70} & \scorecell{3.87} & \scorecell{0.80} & \scorecell{2.73} & \scorecell{1.69} & \scorecell{3.73} \\
MiniMax-Speech-2.7 & ja & 18.1\% & \pairscore{1.79}{2.14} & \pairscore{2.02}{2.36} & \pairscore{1.84}{2.24} & \scorecell{2.17} & \scorecell{2.22} & \scorecell{2.57} & \scorecell{2.51} & \scorecell{1.19} & \scorecell{2.80} & \scorecell{2.20} & \scorecell{2.78} & \scorecell{2.08} & \scorecell{1.40} & \scorecell{1.57} & \scorecell{1.00} & \scorecell{1.00} \\
GPT-4o-Mini-TTS & ja & 4.2\% & \pairscore{1.87}{2.30} & \pairscore{2.11}{2.51} & \pairscore{1.99}{2.49} & \scorecell{1.94} & \scorecell{2.65} & \scorecell{2.48} & \scorecell{2.68} & \scorecell{3.45} & \scorecell{2.20} & \scorecell{2.10} & \scorecell{2.56} & \scorecell{1.71} & \scorecell{2.17} & \scorecell{3.86} & \scorecell{1.80} & \scorecell{1.80} \\

\midrule
\rowcolor{tableblue}
\multicolumn{19}{c}{\normalsize\bfseries Open-source Systems} \\
\midrule
Qwen3TTS-12Hz-1.7B-VD & ja & 3.9\% & \pairscore{2.38}{3.32} & \pairscore{2.63}{3.52} & \pairscore{2.46}{3.55} & \scorecell{3.46} & \scorecell{3.74} & \scorecell{4.13} & \scorecell{3.64} & \scorecell{1.64} & \scorecell{3.50} & \scorecell{3.30} & \scorecell{2.44} & \scorecell{3.10} & \scorecell{1.67} & \scorecell{4.33} & \scorecell{3.00} & \scorecell{1.90} \\
Ming-omni-tts-16.8B-A3B & ja & 57.2\% & \pairscore{0.54}{0.55} & \pairscore{0.34}{0.34} & \pairscore{0.50}{0.51} & \scorecell{0.41} & \scorecell{0.57} & \scorecell{0.53} & \scorecell{0.28} & \scorecell{0.21} & \scorecell{0.40} & \scorecell{0.64} & \scorecell{0.11} & \scorecell{0.30} & \scorecell{1.50} & \scorecell{1.29} & \scorecell{1.10} & \scorecell{1.00} \\
MOSS-VoiceGenerator & ja & 72.9\% & \pairscore{0.35}{0.39} & \pairscore{0.04}{0.04} & \pairscore{0.06}{0.07} & \scorecell{0.15} & \scorecell{0.16} & \scorecell{0.00} & \scorecell{0.00} & \scorecell{0.00} & \scorecell{0.00} & \scorecell{0.00} & \scorecell{0.00} & \scorecell{0.28} & \scorecell{2.17} & \scorecell{3.86} & \scorecell{2.33} & \scorecell{2.11} \\
MiMo-Audio-7B-Instruct & ja & 104.6\% & \pairscore{0.56}{0.59} & \pairscore{0.30}{0.30} & \pairscore{0.44}{0.47} & \scorecell{0.60} & \scorecell{0.42} & \scorecell{0.43} & \scorecell{0.19} & \scorecell{0.42} & \scorecell{0.30} & \scorecell{0.60} & \scorecell{0.00} & \scorecell{0.20} & \scorecell{3.00} & \scorecell{2.86} & \scorecell{1.60} & \scorecell{1.50} \\
Ming-omni-tts-0.5B & ja & 161.7\% & \pairscore{0.16}{0.16} & \pairscore{0.00}{0.00} & \pairscore{0.00}{0.00} & \scorecell{0.00} & \scorecell{0.00} & \scorecell{0.00} & \scorecell{0.00} & \scorecell{0.00} & \scorecell{0.00} & \scorecell{0.00} & \scorecell{0.00} & \scorecell{0.00} & \scorecell{1.00} & \scorecell{1.00} & \scorecell{1.40} & \scorecell{1.00} \\
\bottomrule
\end{tabular}
}
\end{table*}

\begin{table*}[t]
\centering
\small
\setlength{\tabcolsep}{3.8pt}
\renewcommand{\arraystretch}{1.1}

\providecommand{\scorecell}[1]{#1}
\providecommand{\bestscore}[1]{\textbf{#1}}
\providecommand{\secondscore}[1]{\underline{#1}}
\providecommand{\pairscore}[2]{\makecell[c]{#1 / #2}}

\caption{Detailed results on Korean (ko). Scores for Overall, Timbre, and Style are reported as \textit{Instruction-Following (IF) score / Perceptual Quality (PQ) score}. Fine-grained columns report the PQ score.}
\label{tab:main_results_ko}

\resizebox{\textwidth}{!}{
\begin{tabular}{>{\raggedright\arraybackslash\bfseries\normalsize}l l *{17}{c}}
\toprule
\multirow{3}{*}{\normalsize\bfseries Model} 
& \multirow{3}{*}{\normalsize\bfseries Lang} 
& \multirow{3}{*}{\normalsize\bfseries WER $\downarrow$} 
& \multirow{3}{*}{\normalsize\bfseries Overall $\uparrow$} 
& \multirow{3}{*}{\normalsize\bfseries Timbre $\uparrow$} 
& \multirow{3}{*}{\normalsize\bfseries Style $\uparrow$} 
& \multicolumn{3}{c}{\normalsize\bfseries Easy $\uparrow$} 
& \multicolumn{6}{c}{\normalsize\bfseries Hard $\uparrow$} 
& \multicolumn{4}{c}{\normalsize\bfseries Special $\uparrow$} \\
\cmidrule(lr){7-9} \cmidrule(lr){10-15} \cmidrule(lr){16-19}
& & & & & 
& \footnotesize\bfseries Tag & \footnotesize\bfseries Direct & \footnotesize\bfseries B-Comp.
& \footnotesize\bfseries M-Comp. & \footnotesize\bfseries Dynamic & \footnotesize\bfseries Layered & \footnotesize\bfseries Conflict & \footnotesize\bfseries Scenario & \footnotesize\bfseries Character
& \footnotesize\bfseries Disfl. & \footnotesize\bfseries Dysph. & \footnotesize\bfseries Impl. & \footnotesize\bfseries Expl. \\
\midrule

\rowcolor{tableblue}
\multicolumn{19}{c}{\normalsize\bfseries Commercial Systems} \\
\midrule
Gemini 2.5-Flash & ko & 1.4\% & \pairscore{2.65}{4.08} & \pairscore{2.61}{4.00} & \pairscore{2.78}{4.31} & \scorecell{3.92} & \scorecell{4.53} & \scorecell{3.95} & \scorecell{4.77} & \scorecell{3.73} & \scorecell{4.40} & \scorecell{3.10} & \scorecell{4.44} & \scorecell{3.86} & \scorecell{2.00} & \scorecell{3.43} & \scorecell{4.00} & \scorecell{4.20} \\
Gemini 2.5-Pro & ko & 1.2\% & \pairscore{2.68}{4.13} & \pairscore{2.77}{4.26} & \pairscore{2.79}{4.31} & \scorecell{3.68} & \scorecell{4.46} & \scorecell{4.75} & \scorecell{4.59} & \scorecell{4.00} & \scorecell{4.40} & \scorecell{4.30} & \scorecell{4.67} & \scorecell{3.84} & \scorecell{1.17} & \scorecell{4.43} & \scorecell{4.30} & \scorecell{4.20} \\
ElevenLabs-ttv-v3 & ko & 1.2\% & \pairscore{2.49}{3.46} & \pairscore{2.71}{3.57} & \pairscore{2.44}{3.37} & \scorecell{3.50} & \scorecell{3.29} & \scorecell{4.22} & \scorecell{3.57} & \scorecell{1.59} & \scorecell{3.38} & \scorecell{2.70} & \scorecell{4.57} & \scorecell{3.94} & \scorecell{0.85} & \scorecell{3.39} & \scorecell{1.23} & \scorecell{4.42} \\
MiniMax-Speech-2.7 & ko & 1.1\% & \pairscore{2.22}{3.07} & \pairscore{2.58}{3.58} & \pairscore{2.38}{3.33} & \scorecell{3.55} & \scorecell{3.36} & \scorecell{4.26} & \scorecell{3.36} & \scorecell{2.18} & \scorecell{3.30} & \scorecell{3.10} & \scorecell{2.89} & \scorecell{3.05} & \scorecell{1.00} & \scorecell{1.57} & \scorecell{1.44} & \scorecell{1.00} \\
GPT-4o-Mini-TTS & ko & 2.2\% & \pairscore{1.85}{2.33} & \pairscore{1.88}{2.22} & \pairscore{2.02}{2.63} & \scorecell{2.78} & \scorecell{2.18} & \scorecell{3.00} & \scorecell{2.32} & \scorecell{2.55} & \scorecell{2.20} & \scorecell{2.70} & \scorecell{2.00} & \scorecell{1.97} & \scorecell{1.00} & \scorecell{2.29} & \scorecell{1.80} & \scorecell{1.90} \\

\midrule
\rowcolor{tableblue}
\multicolumn{19}{c}{\normalsize\bfseries Open-source Systems} \\
\midrule
Qwen3TTS-12Hz-1.7B-VD & ko & 2.2\% & \pairscore{2.55}{3.77} & \pairscore{2.70}{3.96} & \pairscore{2.60}{3.89} & \scorecell{4.10} & \scorecell{4.19} & \scorecell{4.25} & \scorecell{4.00} & \scorecell{2.73} & \scorecell{3.80} & \scorecell{3.10} & \scorecell{3.25} & \scorecell{3.78} & \scorecell{1.67} & \scorecell{5.00} & \scorecell{2.10} & \scorecell{2.70} \\
Ming-omni-tts-16.8B-A3B & ko & 131.8\% & \pairscore{0.24}{0.25} & \pairscore{0.10}{0.11} & \pairscore{0.09}{0.10} & \scorecell{0.15} & \scorecell{0.07} & \scorecell{0.09} & \scorecell{0.08} & \scorecell{0.00} & \scorecell{0.17} & \scorecell{0.10} & \scorecell{0.00} & \scorecell{0.00} & \scorecell{1.00} & \scorecell{1.00} & \scorecell{1.00} & \scorecell{1.50} \\
MOSS-VoiceGenerator & ko & 66.8\% & \pairscore{0.54}{0.58} & \pairscore{0.37}{0.40} & \pairscore{0.44}{0.49} & \scorecell{0.13} & \scorecell{0.42} & \scorecell{0.92} & \scorecell{0.74} & \scorecell{0.32} & \scorecell{0.24} & \scorecell{0.32} & \scorecell{0.00} & \scorecell{0.31} & \scorecell{1.17} & \scorecell{3.50} & \scorecell{2.00} & \scorecell{1.70} \\
MiMo-Audio-7B-Instruct & ko & 132.1\% & \pairscore{0.46}{0.49} & \pairscore{0.22}{0.22} & \pairscore{0.30}{0.32} & \scorecell{0.40} & \scorecell{0.37} & \scorecell{0.64} & \scorecell{0.00} & \scorecell{0.39} & \scorecell{0.28} & \scorecell{0.00} & \scorecell{0.00} & \scorecell{0.18} & \scorecell{1.50} & \scorecell{3.00} & \scorecell{1.80} & \scorecell{1.70} \\
Ming-omni-tts-0.5B & ko & 207.7\% & \pairscore{0.15}{0.16} & \pairscore{0.00}{0.00} & \pairscore{0.00}{0.00} & \scorecell{0.00} & \scorecell{0.00} & \scorecell{0.00} & \scorecell{0.00} & \scorecell{0.00} & \scorecell{0.00} & \scorecell{0.00} & \scorecell{0.00} & \scorecell{0.00} & \scorecell{1.00} & \scorecell{1.29} & \scorecell{1.00} & \scorecell{1.00} \\
\bottomrule
\end{tabular}
}
\end{table*}

\begin{table*}[t]
\centering
\small
\setlength{\tabcolsep}{3.8pt}
\renewcommand{\arraystretch}{1.1}

\providecommand{\scorecell}[1]{#1}
\providecommand{\bestscore}[1]{\textbf{#1}}
\providecommand{\secondscore}[1]{\underline{#1}}
\providecommand{\pairscore}[2]{\makecell[c]{#1 / #2}}

\caption{Detailed results on French (fr). Scores for Overall, Timbre, and Style are reported as \textit{Instruction-Following (IF) score / Perceptual Quality (PQ) score}. Fine-grained columns report the PQ score.}
\label{tab:main_results_fr}

\resizebox{\textwidth}{!}{
\begin{tabular}{>{\raggedright\arraybackslash\bfseries\normalsize}l l *{17}{c}}
\toprule
\multirow{3}{*}{\normalsize\bfseries Model} 
& \multirow{3}{*}{\normalsize\bfseries Lang} 
& \multirow{3}{*}{\normalsize\bfseries WER $\downarrow$} 
& \multirow{3}{*}{\normalsize\bfseries Overall $\uparrow$} 
& \multirow{3}{*}{\normalsize\bfseries Timbre $\uparrow$} 
& \multirow{3}{*}{\normalsize\bfseries Style $\uparrow$} 
& \multicolumn{3}{c}{\normalsize\bfseries Easy $\uparrow$} 
& \multicolumn{6}{c}{\normalsize\bfseries Hard $\uparrow$} 
& \multicolumn{4}{c}{\normalsize\bfseries Special $\uparrow$} \\
\cmidrule(lr){7-9} \cmidrule(lr){10-15} \cmidrule(lr){16-19}
& & & & & 
& \footnotesize\bfseries Tag & \footnotesize\bfseries Direct & \footnotesize\bfseries B-Comp.
& \footnotesize\bfseries M-Comp. & \footnotesize\bfseries Dynamic & \footnotesize\bfseries Layered & \footnotesize\bfseries Conflict & \footnotesize\bfseries Scenario & \footnotesize\bfseries Character
& \footnotesize\bfseries Disfl. & \footnotesize\bfseries Dysph. & \footnotesize\bfseries Impl. & \footnotesize\bfseries Expl. \\
\midrule

\rowcolor{tableblue}
\multicolumn{19}{c}{\normalsize\bfseries Commercial Systems} \\
\midrule
Gemini 2.5-Flash & fr & 4.3\% & \pairscore{2.46}{3.70} & \pairscore{2.45}{3.57} & \pairscore{2.50}{3.83} & \scorecell{3.53} & \scorecell{3.88} & \scorecell{3.43} & \scorecell{4.73} & \scorecell{3.06} & \scorecell{3.70} & \scorecell{2.72} & \scorecell{4.56} & \scorecell{3.43} & \scorecell{2.33} & \scorecell{5.00} & \scorecell{3.80} & \scorecell{3.90} \\
Gemini 2.5-Pro & fr & 4.1\% & \pairscore{2.51}{3.64} & \pairscore{2.50}{3.52} & \pairscore{2.58}{3.76} & \scorecell{3.30} & \scorecell{3.69} & \scorecell{3.43} & \scorecell{4.27} & \scorecell{4.18} & \scorecell{3.89} & \scorecell{3.33} & \scorecell{4.11} & \scorecell{3.53} & \scorecell{1.17} & \scorecell{5.00} & \scorecell{3.70} & \scorecell{4.20} \\
ElevenLabs-ttv-v3 & fr & 4.1\% & \pairscore{2.30}{3.28} & \pairscore{2.58}{3.60} & \pairscore{2.34}{3.34} & \scorecell{3.66} & \scorecell{3.26} & \scorecell{3.81} & \scorecell{3.55} & \scorecell{2.50} & \scorecell{2.82} & \scorecell{2.15} & \scorecell{4.48} & \scorecell{3.78} & \scorecell{0.80} & \scorecell{2.06} & \scorecell{0.98} & \scorecell{4.13} \\
MiniMax-Speech-2.7 & fr & 4.2\% & \pairscore{2.04}{2.73} & \pairscore{2.24}{2.86} & \pairscore{2.15}{2.93} & \scorecell{2.62} & \scorecell{2.93} & \scorecell{2.83} & \scorecell{3.64} & \scorecell{2.18} & \scorecell{2.50} & \scorecell{3.15} & \scorecell{2.86} & \scorecell{3.35} & \scorecell{1.00} & \scorecell{1.14} & \scorecell{1.00} & \scorecell{1.20} \\
GPT-4o-Mini-TTS & fr & 3.4\% & \pairscore{1.59}{1.77} & \pairscore{1.71}{1.83} & \pairscore{1.64}{1.87} & \scorecell{1.97} & \scorecell{1.91} & \scorecell{1.77} & \scorecell{2.00} & \scorecell{2.27} & \scorecell{1.60} & \scorecell{0.99} & \scorecell{1.62} & \scorecell{1.51} & \scorecell{1.00} & \scorecell{2.00} & \scorecell{1.40} & \scorecell{1.50} \\

\midrule
\rowcolor{tableblue}
\multicolumn{19}{c}{\normalsize\bfseries Open-source Systems} \\
\midrule
Qwen3TTS-12Hz-1.7B-VD & fr & 6.7\% & \pairscore{2.25}{3.17} & \pairscore{2.36}{3.14} & \pairscore{2.28}{3.35} & \scorecell{3.08} & \scorecell{3.29} & \scorecell{3.24} & \scorecell{4.04} & \scorecell{2.27} & \scorecell{2.70} & \scorecell{3.06} & \scorecell{4.11} & \scorecell{3.28} & \scorecell{1.00} & \scorecell{4.43} & \scorecell{1.90} & \scorecell{2.33} \\
Ming-omni-tts-16.8B-A3B & fr & 10.7\% & \pairscore{1.25}{1.26} & \pairscore{1.53}{1.53} & \pairscore{1.26}{1.28} & \scorecell{1.34} & \scorecell{1.34} & \scorecell{1.33} & \scorecell{1.45} & \scorecell{1.49} & \scorecell{0.99} & \scorecell{1.17} & \scorecell{0.86} & \scorecell{1.03} & \scorecell{1.00} & \scorecell{1.00} & \scorecell{1.20} & \scorecell{1.00} \\
MOSS-VoiceGenerator & fr & 54.3\% & \pairscore{0.56}{0.60} & \pairscore{0.44}{0.44} & \pairscore{0.44}{0.48} & \scorecell{0.51} & \scorecell{0.58} & \scorecell{0.65} & \scorecell{0.36} & \scorecell{0.00} & \scorecell{0.26} & \scorecell{0.00} & \scorecell{0.00} & \scorecell{0.28} & \scorecell{1.50} & \scorecell{4.14} & \scorecell{1.40} & \scorecell{1.00} \\
MiMo-Audio-7B-Instruct & fr & 75.6\% & \pairscore{0.31}{0.32} & \pairscore{0.15}{0.15} & \pairscore{0.17}{0.18} & \scorecell{0.13} & \scorecell{0.14} & \scorecell{0.28} & \scorecell{0.16} & \scorecell{0.00} & \scorecell{0.00} & \scorecell{0.00} & \scorecell{0.00} & \scorecell{0.04} & \scorecell{1.50} & \scorecell{2.86} & \scorecell{1.50} & \scorecell{1.30} \\
Ming-omni-tts-0.5B & fr & 44.6\% & \pairscore{0.38}{0.39} & \pairscore{0.37}{0.37} & \pairscore{0.27}{0.27} & \scorecell{0.43} & \scorecell{0.17} & \scorecell{0.18} & \scorecell{0.27} & \scorecell{0.10} & \scorecell{0.20} & \scorecell{0.48} & \scorecell{0.00} & \scorecell{0.14} & \scorecell{1.00} & \scorecell{1.71} & \scorecell{1.11} & \scorecell{1.00} \\
\bottomrule
\end{tabular}
}
\end{table*}

\begin{table*}[t]
\centering
\small
\setlength{\tabcolsep}{3.8pt}
\renewcommand{\arraystretch}{1.1}

\providecommand{\scorecell}[1]{#1}
\providecommand{\bestscore}[1]{\textbf{#1}}
\providecommand{\secondscore}[1]{\underline{#1}}
\providecommand{\pairscore}[2]{\makecell[c]{#1 / #2}}

\caption{Detailed results on German (de). Scores for Overall, Timbre, and Style are reported as \textit{Instruction-Following (IF) score / Perceptual Quality (PQ) score}. Fine-grained columns report the PQ score.}
\label{tab:main_results_de}

\resizebox{\textwidth}{!}{
\begin{tabular}{>{\raggedright\arraybackslash\bfseries\normalsize}l l *{17}{c}}
\toprule
\multirow{3}{*}{\normalsize\bfseries Model} 
& \multirow{3}{*}{\normalsize\bfseries Lang} 
& \multirow{3}{*}{\normalsize\bfseries WER $\downarrow$} 
& \multirow{3}{*}{\normalsize\bfseries Overall $\uparrow$} 
& \multirow{3}{*}{\normalsize\bfseries Timbre $\uparrow$} 
& \multirow{3}{*}{\normalsize\bfseries Style $\uparrow$} 
& \multicolumn{3}{c}{\normalsize\bfseries Easy $\uparrow$} 
& \multicolumn{6}{c}{\normalsize\bfseries Hard $\uparrow$} 
& \multicolumn{4}{c}{\normalsize\bfseries Special $\uparrow$} \\
\cmidrule(lr){7-9} \cmidrule(lr){10-15} \cmidrule(lr){16-19}
& & & & & 
& \footnotesize\bfseries Tag & \footnotesize\bfseries Direct & \footnotesize\bfseries B-Comp.
& \footnotesize\bfseries M-Comp. & \footnotesize\bfseries Dynamic & \footnotesize\bfseries Layered & \footnotesize\bfseries Conflict & \footnotesize\bfseries Scenario & \footnotesize\bfseries Character
& \footnotesize\bfseries Disfl. & \footnotesize\bfseries Dysph. & \footnotesize\bfseries Impl. & \footnotesize\bfseries Expl. \\
\midrule

\rowcolor{tableblue}
\multicolumn{19}{c}{\normalsize\bfseries Commercial Systems} \\
\midrule
Gemini 2.5-Flash & de & 1.6\% & \pairscore{2.60}{3.96} & \pairscore{2.59}{3.87} & \pairscore{2.71}{4.17} & \scorecell{3.88} & \scorecell{4.00} & \scorecell{4.55} & \scorecell{4.27} & \scorecell{3.91} & \scorecell{3.56} & \scorecell{3.89} & \scorecell{5.00} & \scorecell{3.97} & \scorecell{1.83} & \scorecell{4.43} & \scorecell{3.30} & \scorecell{3.30} \\
Gemini 2.5-Pro & de & 1.8\% & \pairscore{2.49}{3.72} & \pairscore{2.50}{3.62} & \pairscore{2.58}{3.87} & \scorecell{3.62} & \scorecell{3.58} & \scorecell{4.14} & \scorecell{3.99} & \scorecell{4.18} & \scorecell{4.00} & \scorecell{3.42} & \scorecell{4.78} & \scorecell{3.60} & \scorecell{1.83} & \scorecell{3.29} & \scorecell{3.60} & \scorecell{3.90} \\
ElevenLabs-ttv-v3 & de & 2.5\% & \pairscore{2.41}{3.52} & \pairscore{2.70}{3.78} & \pairscore{2.42}{3.51} & \scorecell{3.81} & \scorecell{3.65} & \scorecell{3.99} & \scorecell{3.45} & \scorecell{2.00} & \scorecell{3.41} & \scorecell{3.44} & \scorecell{4.37} & \scorecell{4.12} & \scorecell{0.81} & \scorecell{2.28} & \scorecell{2.00} & \scorecell{3.40} \\
MiniMax-Speech-2.7 & de & 3.6\% & \pairscore{2.05}{2.74} & \pairscore{2.26}{2.95} & \pairscore{2.10}{2.89} & \scorecell{2.97} & \scorecell{2.76} & \scorecell{3.68} & \scorecell{3.18} & \scorecell{1.56} & \scorecell{3.20} & \scorecell{2.25} & \scorecell{4.33} & \scorecell{2.97} & \scorecell{1.00} & \scorecell{1.00} & \scorecell{1.50} & \scorecell{1.00} \\
GPT-4o-Mini-TTS & de & 2.1\% & \pairscore{1.53}{1.77} & \pairscore{1.87}{2.14} & \pairscore{1.46}{1.69} & \scorecell{2.17} & \scorecell{1.96} & \scorecell{2.14} & \scorecell{1.52} & \scorecell{1.82} & \scorecell{1.26} & \scorecell{1.20} & \scorecell{2.33} & \scorecell{1.56} & \scorecell{1.67} & \scorecell{1.00} & \scorecell{1.20} & \scorecell{1.00} \\

\midrule
\rowcolor{tableblue}
\multicolumn{19}{c}{\normalsize\bfseries Open-source Systems} \\
\midrule
Qwen3TTS-12Hz-1.7B-VD & de & 2.8\% & \pairscore{2.47}{3.57} & \pairscore{2.60}{3.71} & \pairscore{2.53}{3.74} & \scorecell{3.81} & \scorecell{3.45} & \scorecell{4.50} & \scorecell{3.64} & \scorecell{2.56} & \scorecell{4.20} & \scorecell{4.30} & \scorecell{3.89} & \scorecell{3.53} & \scorecell{1.67} & \scorecell{3.00} & \scorecell{3.10} & \scorecell{2.40} \\
Ming-omni-tts-16.8B-A3B & de & 10.7\% & \pairscore{1.32}{1.39} & \pairscore{1.49}{1.52} & \pairscore{1.25}{1.32} & \scorecell{1.61} & \scorecell{1.34} & \scorecell{2.04} & \scorecell{0.98} & \scorecell{1.18} & \scorecell{1.30} & \scorecell{0.84} & \scorecell{0.93} & \scorecell{1.18} & \scorecell{1.40} & \scorecell{1.57} & \scorecell{1.60} & \scorecell{1.20} \\
MOSS-VoiceGenerator & de & 41.7\% & \pairscore{0.61}{0.67} & \pairscore{0.48}{0.50} & \pairscore{0.42}{0.48} & \scorecell{0.56} & \scorecell{0.50} & \scorecell{0.43} & \scorecell{0.50} & \scorecell{0.50} & \scorecell{0.12} & \scorecell{0.45} & \scorecell{0.32} & \scorecell{0.41} & \scorecell{1.17} & \scorecell{2.86} & \scorecell{1.60} & \scorecell{1.00} \\
MiMo-Audio-7B-Instruct & de & 74.1\% & \pairscore{0.27}{0.28} & \pairscore{0.08}{0.08} & \pairscore{0.03}{0.03} & \scorecell{0.13} & \scorecell{0.08} & \scorecell{0.00} & \scorecell{0.00} & \scorecell{0.00} & \scorecell{0.00} & \scorecell{0.00} & \scorecell{0.00} & \scorecell{0.05} & \scorecell{1.17} & \scorecell{1.71} & \scorecell{2.30} & \scorecell{1.30} \\
Ming-omni-tts-0.5B & de & 53.4\% & \pairscore{0.25}{0.26} & \pairscore{0.13}{0.13} & \pairscore{0.11}{0.11} & \scorecell{0.07} & \scorecell{0.20} & \scorecell{0.24} & \scorecell{0.00} & \scorecell{0.00} & \scorecell{0.14} & \scorecell{0.12} & \scorecell{0.00} & \scorecell{0.00} & \scorecell{1.33} & \scorecell{1.43} & \scorecell{1.22} & \scorecell{1.00} \\
\bottomrule
\end{tabular}
}
\end{table*}

\begin{table*}[t]
\centering
\small
\setlength{\tabcolsep}{3.8pt}
\renewcommand{\arraystretch}{1.1}

\providecommand{\scorecell}[1]{#1}
\providecommand{\bestscore}[1]{\textbf{#1}}
\providecommand{\secondscore}[1]{\underline{#1}}
\providecommand{\pairscore}[2]{\makecell[c]{#1 / #2}}

\caption{Detailed results on Spanish (es). Scores for Overall, Timbre, and Style are reported as \textit{Instruction-Following (IF) score / Perceptual Quality (PQ) score}. Fine-grained columns report the PQ score.}
\label{tab:main_results_es}

\resizebox{\textwidth}{!}{
\begin{tabular}{>{\raggedright\arraybackslash\bfseries\normalsize}l l *{17}{c}}
\toprule
\multirow{3}{*}{\normalsize\bfseries Model} 
& \multirow{3}{*}{\normalsize\bfseries Lang} 
& \multirow{3}{*}{\normalsize\bfseries WER $\downarrow$} 
& \multirow{3}{*}{\normalsize\bfseries Overall $\uparrow$} 
& \multirow{3}{*}{\normalsize\bfseries Timbre $\uparrow$} 
& \multirow{3}{*}{\normalsize\bfseries Style $\uparrow$} 
& \multicolumn{3}{c}{\normalsize\bfseries Easy $\uparrow$} 
& \multicolumn{6}{c}{\normalsize\bfseries Hard $\uparrow$} 
& \multicolumn{4}{c}{\normalsize\bfseries Special $\uparrow$} \\
\cmidrule(lr){7-9} \cmidrule(lr){10-15} \cmidrule(lr){16-19}
& & & & & 
& \footnotesize\bfseries Tag & \footnotesize\bfseries Direct & \footnotesize\bfseries B-Comp.
& \footnotesize\bfseries M-Comp. & \footnotesize\bfseries Dynamic & \footnotesize\bfseries Layered & \footnotesize\bfseries Conflict & \footnotesize\bfseries Scenario & \footnotesize\bfseries Character
& \footnotesize\bfseries Disfl. & \footnotesize\bfseries Dysph. & \footnotesize\bfseries Impl. & \footnotesize\bfseries Expl. \\
\midrule

\rowcolor{tableblue}
\multicolumn{19}{c}{\normalsize\bfseries Commercial Systems} \\
\midrule
Gemini 2.5-Flash & es & 1.9\% & \pairscore{2.51}{3.70} & \pairscore{2.40}{3.36} & \pairscore{2.54}{3.80} & \scorecell{3.49} & \scorecell{3.98} & \scorecell{3.72} & \scorecell{3.42} & \scorecell{5.00} & \scorecell{3.24} & \scorecell{2.88} & \scorecell{4.11} & \scorecell{3.68} & \scorecell{1.83} & \scorecell{4.71} & \scorecell{4.20} & \scorecell{3.50} \\
Gemini 2.5-Pro & es & 1.9\% & \pairscore{2.42}{3.49} & \pairscore{2.45}{3.40} & \pairscore{2.41}{3.49} & \scorecell{3.53} & \scorecell{3.43} & \scorecell{3.53} & \scorecell{3.86} & \scorecell{4.18} & \scorecell{2.61} & \scorecell{2.48} & \scorecell{3.56} & \scorecell{2.99} & \scorecell{2.00} & \scorecell{4.43} & \scorecell{4.60} & \scorecell{4.70} \\
ElevenLabs-ttv-v3 & es & 1.7\% & \pairscore{2.34}{3.25} & \pairscore{2.58}{3.46} & \pairscore{2.35}{3.35} & \scorecell{3.62} & \scorecell{3.31} & \scorecell{3.45} & \scorecell{4.23} & \scorecell{2.08} & \scorecell{2.52} & \scorecell{2.20} & \scorecell{3.55} & \scorecell{3.85} & \scorecell{0.81} & \scorecell{2.65} & \scorecell{1.37} & \scorecell{2.94} \\
MiniMax-Speech-2.7 & es & 1.6\% & \pairscore{2.10}{2.87} & \pairscore{2.32}{3.16} & \pairscore{2.15}{3.03} & \scorecell{2.95} & \scorecell{3.04} & \scorecell{3.17} & \scorecell{3.77} & \scorecell{1.73} & \scorecell{4.10} & \scorecell{2.56} & \scorecell{3.16} & \scorecell{3.10} & \scorecell{1.00} & \scorecell{2.14} & \scorecell{1.40} & \scorecell{1.00} \\
GPT-4o-Mini-TTS & es & 2.1\% & \pairscore{1.49}{1.68} & \pairscore{1.62}{1.76} & \pairscore{1.51}{1.67} & \scorecell{1.73} & \scorecell{2.11} & \scorecell{1.96} & \scorecell{1.57} & \scorecell{1.27} & \scorecell{1.53} & \scorecell{1.08} & \scorecell{1.73} & \scorecell{1.22} & \scorecell{1.00} & \scorecell{1.57} & \scorecell{2.33} & \scorecell{1.60} \\

\midrule
\rowcolor{tableblue}
\multicolumn{19}{c}{\normalsize\bfseries Open-source Systems} \\
\midrule
Qwen3TTS-12Hz-1.7B-VD & es & 2.5\% & \pairscore{2.29}{3.29} & \pairscore{2.45}{3.40} & \pairscore{2.30}{3.38} & \scorecell{3.20} & \scorecell{3.98} & \scorecell{3.49} & \scorecell{3.30} & \scorecell{2.64} & \scorecell{2.32} & \scorecell{3.04} & \scorecell{2.07} & \scorecell{3.69} & \scorecell{1.00} & \scorecell{4.57} & \scorecell{2.00} & \scorecell{3.00} \\
Ming-omni-tts-16.8B-A3B & es & 5.5\% & \pairscore{1.28}{1.30} & \pairscore{1.62}{1.65} & \pairscore{1.28}{1.32} & \scorecell{1.76} & \scorecell{1.23} & \scorecell{1.52} & \scorecell{1.49} & \scorecell{1.19} & \scorecell{0.96} & \scorecell{0.99} & \scorecell{0.95} & \scorecell{0.96} & \scorecell{1.00} & \scorecell{1.14} & \scorecell{1.10} & \scorecell{1.00} \\
MOSS-VoiceGenerator & es & 15.5\% & \pairscore{1.09}{1.20} & \pairscore{1.07}{1.11} & \pairscore{1.00}{1.09} & \scorecell{0.93} & \scorecell{1.18} & \scorecell{0.86} & \scorecell{1.30} & \scorecell{0.90} & \scorecell{0.95} & \scorecell{1.47} & \scorecell{0.59} & \scorecell{1.01} & \scorecell{1.50} & \scorecell{4.00} & \scorecell{2.30} & \scorecell{1.10} \\
MiMo-Audio-7B-Instruct & es & 44.2\% & \pairscore{0.34}{0.35} & \pairscore{0.27}{0.27} & \pairscore{0.14}{0.14} & \scorecell{0.28} & \scorecell{0.28} & \scorecell{0.16} & \scorecell{0.00} & \scorecell{0.25} & \scorecell{0.10} & \scorecell{0.12} & \scorecell{0.00} & \scorecell{0.04} & \scorecell{1.00} & \scorecell{2.57} & \scorecell{1.00} & \scorecell{1.00} \\
Ming-omni-tts-0.5B & es & 39.7\% & \pairscore{0.30}{0.31} & \pairscore{0.28}{0.28} & \pairscore{0.11}{0.11} & \scorecell{0.22} & \scorecell{0.18} & \scorecell{0.29} & \scorecell{0.14} & \scorecell{0.00} & \scorecell{0.00} & \scorecell{0.10} & \scorecell{0.11} & \scorecell{0.06} & \scorecell{1.00} & \scorecell{1.29} & \scorecell{1.00} & \scorecell{1.00} \\
\bottomrule
\end{tabular}
}
\end{table*}

\begin{table*}[t]
\centering
\small
\setlength{\tabcolsep}{3.8pt}
\renewcommand{\arraystretch}{1.1}

\providecommand{\scorecell}[1]{#1}
\providecommand{\bestscore}[1]{\textbf{#1}}
\providecommand{\secondscore}[1]{\underline{#1}}
\providecommand{\pairscore}[2]{\makecell[c]{#1 / #2}}

\caption{Detailed results on Portuguese (pt). Scores for Overall, Timbre, and Style are reported as \textit{Instruction-Following (IF) score / Perceptual Quality (PQ) score}. Fine-grained columns report the PQ score.}
\label{tab:main_results_pt}

\resizebox{\textwidth}{!}{
\begin{tabular}{>{\raggedright\arraybackslash\bfseries\normalsize}l l *{17}{c}}
\toprule
\multirow{3}{*}{\normalsize\bfseries Model} 
& \multirow{3}{*}{\normalsize\bfseries Lang} 
& \multirow{3}{*}{\normalsize\bfseries WER $\downarrow$} 
& \multirow{3}{*}{\normalsize\bfseries Overall $\uparrow$} 
& \multirow{3}{*}{\normalsize\bfseries Timbre $\uparrow$} 
& \multirow{3}{*}{\normalsize\bfseries Style $\uparrow$} 
& \multicolumn{3}{c}{\normalsize\bfseries Easy $\uparrow$} 
& \multicolumn{6}{c}{\normalsize\bfseries Hard $\uparrow$} 
& \multicolumn{4}{c}{\normalsize\bfseries Special $\uparrow$} \\
\cmidrule(lr){7-9} \cmidrule(lr){10-15} \cmidrule(lr){16-19}
& & & & & 
& \footnotesize\bfseries Tag & \footnotesize\bfseries Direct & \footnotesize\bfseries B-Comp.
& \footnotesize\bfseries M-Comp. & \footnotesize\bfseries Dynamic & \footnotesize\bfseries Layered & \footnotesize\bfseries Conflict & \footnotesize\bfseries Scenario & \footnotesize\bfseries Character
& \footnotesize\bfseries Disfl. & \footnotesize\bfseries Dysph. & \footnotesize\bfseries Impl. & \footnotesize\bfseries Expl. \\
\midrule

\rowcolor{tableblue}
\multicolumn{19}{c}{\normalsize\bfseries Commercial Systems} \\
\midrule
Gemini 2.5-Flash & pt & 2.4\% & \pairscore{2.54}{3.80} & \pairscore{2.53}{3.65} & \pairscore{2.66}{4.05} & \scorecell{3.53} & \scorecell{4.09} & \scorecell{4.09} & \scorecell{4.27} & \scorecell{2.98} & \scorecell{3.80} & \scorecell{3.67} & \scorecell{4.12} & \scorecell{3.58} & \scorecell{2.67} & \scorecell{4.14} & \scorecell{3.90} & \scorecell{3.80} \\
Gemini 2.5-Pro & pt & 2.9\% & \pairscore{2.48}{3.64} & \pairscore{2.57}{3.60} & \pairscore{2.58}{3.83} & \scorecell{3.48} & \scorecell{3.71} & \scorecell{4.14} & \scorecell{4.41} & \scorecell{3.06} & \scorecell{3.30} & \scorecell{3.50} & \scorecell{3.38} & \scorecell{3.43} & \scorecell{1.67} & \scorecell{3.57} & \scorecell{4.10} & \scorecell{4.20} \\
ElevenLabs-ttv-v3 & pt & 2.6\% & \pairscore{2.33}{3.24} & \pairscore{2.64}{3.55} & \pairscore{2.29}{3.23} & \scorecell{3.15} & \scorecell{3.69} & \scorecell{3.91} & \scorecell{3.41} & \scorecell{2.41} & \scorecell{2.53} & \scorecell{2.40} & \scorecell{3.65} & \scorecell{3.84} & \scorecell{0.82} & \scorecell{1.69} & \scorecell{1.28} & \scorecell{4.15} \\
MiniMax-Speech-2.7 & pt & 2.4\% & \pairscore{2.12}{2.94} & \pairscore{2.28}{3.22} & \pairscore{2.24}{3.18} & \scorecell{3.02} & \scorecell{3.12} & \scorecell{3.82} & \scorecell{3.68} & \scorecell{2.36} & \scorecell{2.50} & \scorecell{2.60} & \scorecell{3.62} & \scorecell{3.00} & \scorecell{1.00} & \scorecell{1.57} & \scorecell{2.20} & \scorecell{1.00} \\
GPT-4o-Mini-TTS & pt & 3.1\% & \pairscore{1.72}{2.15} & \pairscore{1.76}{2.08} & \pairscore{1.73}{2.23} & \scorecell{1.94} & \scorecell{2.33} & \scorecell{2.47} & \scorecell{2.00} & \scorecell{2.73} & \scorecell{2.16} & \scorecell{1.80} & \scorecell{2.88} & \scorecell{1.79} & \scorecell{1.67} & \scorecell{3.00} & \scorecell{1.90} & \scorecell{2.30} \\

\midrule
\rowcolor{tableblue}
\multicolumn{19}{c}{\normalsize\bfseries Open-source Systems} \\
\midrule
Qwen3TTS-12Hz-1.7B-VD & pt & 2.8\% & \pairscore{2.37}{3.21} & \pairscore{2.62}{3.39} & \pairscore{2.46}{3.44} & \scorecell{3.24} & \scorecell{3.24} & \scorecell{3.90} & \scorecell{3.91} & \scorecell{2.27} & \scorecell{2.50} & \scorecell{3.56} & \scorecell{2.62} & \scorecell{3.02} & \scorecell{1.17} & \scorecell{4.43} & \scorecell{2.60} & \scorecell{3.40} \\
Ming-omni-tts-16.8B-A3B & pt & 11.9\% & \pairscore{1.10}{1.10} & \pairscore{1.32}{1.32} & \pairscore{1.08}{1.08} & \scorecell{1.17} & \scorecell{1.11} & \scorecell{1.37} & \scorecell{1.23} & \scorecell{0.86} & \scorecell{1.20} & \scorecell{1.08} & \scorecell{0.75} & \scorecell{0.86} & \scorecell{1.00} & \scorecell{1.00} & \scorecell{1.00} & \scorecell{1.00} \\
MOSS-VoiceGenerator & pt & 37.5\% & \pairscore{0.61}{0.65} & \pairscore{0.57}{0.58} & \pairscore{0.53}{0.58} & \scorecell{0.55} & \scorecell{0.64} & \scorecell{1.12} & \scorecell{0.33} & \scorecell{0.35} & \scorecell{0.36} & \scorecell{0.45} & \scorecell{0.19} & \scorecell{0.13} & \scorecell{1.33} & \scorecell{3.57} & \scorecell{2.00} & \scorecell{1.20} \\
MiMo-Audio-7B-Instruct & pt & 104.8\% & \pairscore{0.19}{0.19} & \pairscore{0.02}{0.02} & \pairscore{0.00}{0.00} & \scorecell{0.04} & \scorecell{0.00} & \scorecell{0.00} & \scorecell{0.00} & \scorecell{0.00} & \scorecell{0.00} & \scorecell{0.00} & \scorecell{0.00} & \scorecell{0.00} & \scorecell{1.00} & \scorecell{1.14} & \scorecell{1.20} & \scorecell{1.00} \\
Ming-omni-tts-0.5B & pt & 50.6\% & \pairscore{0.29}{0.29} & \pairscore{0.24}{0.24} & \pairscore{0.11}{0.11} & \scorecell{0.16} & \scorecell{0.18} & \scorecell{0.17} & \scorecell{0.13} & \scorecell{0.00} & \scorecell{0.10} & \scorecell{0.11} & \scorecell{0.25} & \scorecell{0.03} & \scorecell{1.33} & \scorecell{1.86} & \scorecell{1.00} & \scorecell{1.00} \\
\bottomrule
\end{tabular}
}
\end{table*}

\begin{table*}[t]
\centering
\small
\setlength{\tabcolsep}{3.8pt}
\renewcommand{\arraystretch}{1.1}

\providecommand{\scorecell}[1]{#1}
\providecommand{\bestscore}[1]{\textbf{#1}}
\providecommand{\secondscore}[1]{\underline{#1}}
\providecommand{\pairscore}[2]{\makecell[c]{#1 / #2}}

\caption{Detailed results on Italian (it). Scores for Overall, Timbre, and Style are reported as \textit{Instruction-Following (IF) score / Perceptual Quality (PQ) score}. Fine-grained columns report the PQ score.}
\label{tab:main_results_it}

\resizebox{\textwidth}{!}{
\begin{tabular}{>{\raggedright\arraybackslash\bfseries\normalsize}l l *{17}{c}}
\toprule
\multirow{3}{*}{\normalsize\bfseries Model} 
& \multirow{3}{*}{\normalsize\bfseries Lang} 
& \multirow{3}{*}{\normalsize\bfseries WER $\downarrow$} 
& \multirow{3}{*}{\normalsize\bfseries Overall $\uparrow$} 
& \multirow{3}{*}{\normalsize\bfseries Timbre $\uparrow$} 
& \multirow{3}{*}{\normalsize\bfseries Style $\uparrow$} 
& \multicolumn{3}{c}{\normalsize\bfseries Easy $\uparrow$} 
& \multicolumn{6}{c}{\normalsize\bfseries Hard $\uparrow$} 
& \multicolumn{4}{c}{\normalsize\bfseries Special $\uparrow$} \\
\cmidrule(lr){7-9} \cmidrule(lr){10-15} \cmidrule(lr){16-19}
& & & & & 
& \footnotesize\bfseries Tag & \footnotesize\bfseries Direct & \footnotesize\bfseries B-Comp.
& \footnotesize\bfseries M-Comp. & \footnotesize\bfseries Dynamic & \footnotesize\bfseries Layered & \footnotesize\bfseries Conflict & \footnotesize\bfseries Scenario & \footnotesize\bfseries Character
& \footnotesize\bfseries Disfl. & \footnotesize\bfseries Dysph. & \footnotesize\bfseries Impl. & \footnotesize\bfseries Expl. \\
\midrule

\rowcolor{tableblue}
\multicolumn{19}{c}{\normalsize\bfseries Commercial Systems} \\
\midrule
Gemini 2.5-Flash & it & 2.6\% & \pairscore{2.49}{3.72} & \pairscore{2.41}{3.47} & \pairscore{2.62}{4.02} & \scorecell{3.90} & \scorecell{3.57} & \scorecell{4.00} & \scorecell{3.99} & \scorecell{4.55} & \scorecell{3.30} & \scorecell{3.15} & \scorecell{4.44} & \scorecell{3.47} & \scorecell{1.67} & \scorecell{4.67} & \scorecell{3.11} & \scorecell{3.70} \\
Gemini 2.5-Pro & it & 2.3\% & \pairscore{2.42}{3.51} & \pairscore{2.47}{3.48} & \pairscore{2.53}{3.76} & \scorecell{3.48} & \scorecell{3.50} & \scorecell{4.09} & \scorecell{3.91} & \scorecell{3.97} & \scorecell{3.10} & \scorecell{3.24} & \scorecell{3.44} & \scorecell{3.10} & \scorecell{2.17} & \scorecell{4.43} & \scorecell{3.30} & \scorecell{3.50} \\
ElevenLabs-ttv-v3 & it & 2.2\% & \pairscore{2.20}{3.00} & \pairscore{2.60}{3.52} & \pairscore{2.13}{2.86} & \scorecell{3.16} & \scorecell{3.14} & \scorecell{3.90} & \scorecell{3.45} & \scorecell{1.76} & \scorecell{1.89} & \scorecell{2.21} & \scorecell{3.41} & \scorecell{3.61} & \scorecell{0.82} & \scorecell{2.29} & \scorecell{1.64} & \scorecell{2.76} \\
MiniMax-Speech-2.7 & it & 2.1\% & \pairscore{2.02}{2.71} & \pairscore{2.28}{3.05} & \pairscore{2.15}{2.96} & \scorecell{2.72} & \scorecell{3.12} & \scorecell{3.59} & \scorecell{3.43} & \scorecell{2.18} & \scorecell{2.10} & \scorecell{2.43} & \scorecell{3.33} & \scorecell{2.65} & \scorecell{1.00} & \scorecell{1.57} & \scorecell{1.80} & \scorecell{1.00} \\
GPT-4o-Mini-TTS & it & 2.4\% & \pairscore{1.46}{1.60} & \pairscore{1.59}{1.63} & \pairscore{1.49}{1.67} & \scorecell{1.71} & \scorecell{1.58} & \scorecell{1.78} & \scorecell{1.26} & \scorecell{2.60} & \scorecell{1.20} & \scorecell{2.10} & \scorecell{1.88} & \scorecell{1.25} & \scorecell{1.00} & \scorecell{2.71} & \scorecell{1.40} & \scorecell{1.20} \\

\midrule
\rowcolor{tableblue}
\multicolumn{19}{c}{\normalsize\bfseries Open-source Systems} \\
\midrule
Qwen3TTS-12Hz-1.7B-VD & it & 3.3\% & \pairscore{2.23}{2.94} & \pairscore{2.44}{3.18} & \pairscore{2.33}{3.13} & \scorecell{2.69} & \scorecell{3.22} & \scorecell{4.14} & \scorecell{3.91} & \scorecell{2.23} & \scorecell{2.50} & \scorecell{1.98} & \scorecell{2.44} & \scorecell{3.03} & \scorecell{2.00} & \scorecell{3.86} & \scorecell{1.60} & \scorecell{1.30} \\
Ming-omni-tts-16.8B-A3B & it & 6.9\% & \pairscore{1.19}{1.21} & \pairscore{1.34}{1.34} & \pairscore{1.21}{1.23} & \scorecell{1.31} & \scorecell{1.23} & \scorecell{1.81} & \scorecell{1.15} & \scorecell{1.19} & \scorecell{0.90} & \scorecell{0.84} & \scorecell{0.74} & \scorecell{1.03} & \scorecell{1.00} & \scorecell{1.29} & \scorecell{1.00} & \scorecell{1.00} \\
MOSS-VoiceGenerator & it & 25.2\% & \pairscore{0.82}{0.89} & \pairscore{0.89}{0.92} & \pairscore{0.77}{0.85} & \scorecell{0.96} & \scorecell{0.79} & \scorecell{1.07} & \scorecell{0.76} & \scorecell{0.53} & \scorecell{0.48} & \scorecell{0.90} & \scorecell{0.11} & \scorecell{0.42} & \scorecell{1.40} & \scorecell{4.00} & \scorecell{2.10} & \scorecell{1.00} \\
Ming-omni-tts-0.5B & it & 34.3\% & \pairscore{0.39}{0.39} & \pairscore{0.47}{0.47} & \pairscore{0.22}{0.23} & \scorecell{0.44} & \scorecell{0.39} & \scorecell{0.45} & \scorecell{0.26} & \scorecell{0.00} & \scorecell{0.00} & \scorecell{0.12} & \scorecell{0.00} & \scorecell{0.03} & \scorecell{1.33} & \scorecell{1.57} & \scorecell{1.10} & \scorecell{1.00} \\
MiMo-Audio-7B-Instruct & it & 47.3\% & \pairscore{0.34}{0.35} & \pairscore{0.18}{0.18} & \pairscore{0.10}{0.11} & \scorecell{0.12} & \scorecell{0.12} & \scorecell{0.21} & \scorecell{0.19} & \scorecell{0.12} & \scorecell{0.10} & \scorecell{0.00} & \scorecell{0.12} & \scorecell{0.15} & \scorecell{2.00} & \scorecell{1.86} & \scorecell{1.20} & \scorecell{1.00} \\
\bottomrule
\end{tabular}
}
\end{table*}

\begin{table*}[t]
\centering
\small
\setlength{\tabcolsep}{3.8pt}
\renewcommand{\arraystretch}{1.1}

\providecommand{\scorecell}[1]{#1}
\providecommand{\bestscore}[1]{\textbf{#1}}
\providecommand{\secondscore}[1]{\underline{#1}}
\providecommand{\pairscore}[2]{\makecell[c]{#1 / #2}}

\caption{Detailed results on Russian (ru). Scores for Overall, Timbre, and Style are reported as \textit{Instruction-Following (IF) score / Perceptual Quality (PQ) score}. Fine-grained columns report the PQ score.}
\label{tab:main_results_ru}

\resizebox{\textwidth}{!}{
\begin{tabular}{>{\raggedright\arraybackslash\bfseries\normalsize}l l *{17}{c}}
\toprule
\multirow{3}{*}{\normalsize\bfseries Model} 
& \multirow{3}{*}{\normalsize\bfseries Lang} 
& \multirow{3}{*}{\normalsize\bfseries WER $\downarrow$} 
& \multirow{3}{*}{\normalsize\bfseries Overall $\uparrow$} 
& \multirow{3}{*}{\normalsize\bfseries Timbre $\uparrow$} 
& \multirow{3}{*}{\normalsize\bfseries Style $\uparrow$} 
& \multicolumn{3}{c}{\normalsize\bfseries Easy $\uparrow$} 
& \multicolumn{6}{c}{\normalsize\bfseries Hard $\uparrow$} 
& \multicolumn{4}{c}{\normalsize\bfseries Special $\uparrow$} \\
\cmidrule(lr){7-9} \cmidrule(lr){10-15} \cmidrule(lr){16-19}
& & & & & 
& \footnotesize\bfseries Tag & \footnotesize\bfseries Direct & \footnotesize\bfseries B-Comp.
& \footnotesize\bfseries M-Comp. & \footnotesize\bfseries Dynamic & \footnotesize\bfseries Layered & \footnotesize\bfseries Conflict & \footnotesize\bfseries Scenario & \footnotesize\bfseries Character
& \footnotesize\bfseries Disfl. & \footnotesize\bfseries Dysph. & \footnotesize\bfseries Impl. & \footnotesize\bfseries Expl. \\
\midrule

\rowcolor{tableblue}
\multicolumn{19}{c}{\normalsize\bfseries Commercial Systems} \\
\midrule
Gemini 2.5-Flash & ru & 5.8\% & \pairscore{2.59}{3.94} & \pairscore{2.47}{3.65} & \pairscore{2.63}{4.01} & \scorecell{3.71} & \scorecell{4.10} & \scorecell{4.43} & \scorecell{3.68} & \scorecell{3.91} & \scorecell{3.90} & \scorecell{3.44} & \scorecell{4.22} & \scorecell{4.16} & \scorecell{2.17} & \scorecell{4.14} & \scorecell{4.30} & \scorecell{3.90} \\
Gemini 2.5-Pro & ru & 6.6\% & \pairscore{2.62}{3.80} & \pairscore{2.57}{3.55} & \pairscore{2.69}{3.95} & \scorecell{3.55} & \scorecell{3.55} & \scorecell{4.18} & \scorecell{4.18} & \scorecell{3.90} & \scorecell{4.00} & \scorecell{3.22} & \scorecell{4.33} & \scorecell{3.89} & \scorecell{2.50} & \scorecell{5.00} & \scorecell{4.00} & \scorecell{3.70} \\
ElevenLabs-ttv-v3 & ru & 6.3\% & \pairscore{2.37}{3.29} & \pairscore{2.69}{3.63} & \pairscore{2.37}{3.19} & \scorecell{3.29} & \scorecell{3.33} & \scorecell{3.90} & \scorecell{3.57} & \scorecell{2.78} & \scorecell{2.66} & \scorecell{2.20} & \scorecell{4.05} & \scorecell{4.16} & \scorecell{0.83} & \scorecell{1.84} & \scorecell{1.02} & \scorecell{3.57} \\
MiniMax-Speech-2.7 & ru & 6.0\% & \pairscore{2.03}{2.61} & \pairscore{2.33}{3.05} & \pairscore{2.16}{2.87} & \scorecell{2.90} & \scorecell{2.55} & \scorecell{3.77} & \scorecell{3.32} & \scorecell{2.73} & \scorecell{2.40} & \scorecell{2.00} & \scorecell{2.67} & \scorecell{2.54} & \scorecell{1.80} & \scorecell{1.57} & \scorecell{1.00} & \scorecell{1.00} \\
GPT-4o-Mini-TTS & ru & 5.8\% & \pairscore{1.88}{2.35} & \pairscore{1.96}{2.36} & \pairscore{2.05}{2.66} & \scorecell{2.44} & \scorecell{2.56} & \scorecell{2.90} & \scorecell{2.77} & \scorecell{2.48} & \scorecell{2.20} & \scorecell{2.22} & \scorecell{1.89} & \scorecell{1.79} & \scorecell{1.17} & \scorecell{3.00} & \scorecell{1.50} & \scorecell{2.50} \\

\midrule
\rowcolor{tableblue}
\multicolumn{19}{c}{\normalsize\bfseries Open-source Systems} \\
\midrule
Qwen3TTS-12Hz-1.7B-VD & ru & 10.0\% & \pairscore{2.05}{2.63} & \pairscore{2.27}{2.83} & \pairscore{2.08}{2.67} & \scorecell{2.25} & \scorecell{2.86} & \scorecell{3.82} & \scorecell{3.09} & \scorecell{2.36} & \scorecell{2.89} & \scorecell{1.68} & \scorecell{1.22} & \scorecell{2.84} & \scorecell{1.67} & \scorecell{2.57} & \scorecell{1.90} & \scorecell{2.60} \\
Ming-omni-tts-16.8B-A3B & ru & 39.9\% & \pairscore{0.89}{0.91} & \pairscore{1.15}{1.16} & \pairscore{0.77}{0.78} & \scorecell{1.09} & \scorecell{0.71} & \scorecell{0.82} & \scorecell{0.99} & \scorecell{0.99} & \scorecell{0.80} & \scorecell{0.68} & \scorecell{0.56} & \scorecell{0.70} & \scorecell{1.00} & \scorecell{1.57} & \scorecell{1.00} & \scorecell{1.30} \\
MOSS-VoiceGenerator & ru & 86.7\% & \pairscore{0.30}{0.31} & \pairscore{0.08}{0.08} & \pairscore{0.12}{0.13} & \scorecell{0.16} & \scorecell{0.00} & \scorecell{0.21} & \scorecell{0.17} & \scorecell{0.00} & \scorecell{0.00} & \scorecell{0.16} & \scorecell{0.00} & \scorecell{0.00} & \scorecell{1.67} & \scorecell{2.86} & \scorecell{1.60} & \scorecell{1.50} \\
MiMo-Audio-7B-Instruct & ru & 104.8\% & \pairscore{0.19}{0.19} & \pairscore{0.02}{0.02} & \pairscore{0.00}{0.00} & \scorecell{0.04} & \scorecell{0.00} & \scorecell{0.00} & \scorecell{0.00} & \scorecell{0.00} & \scorecell{0.00} & \scorecell{0.00} & \scorecell{0.00} & \scorecell{0.00} & \scorecell{1.00} & \scorecell{1.14} & \scorecell{1.20} & \scorecell{1.00} \\
Ming-omni-tts-0.5B & ru & 179.3\% & \pairscore{0.16}{0.17} & \pairscore{0.00}{0.00} & \pairscore{0.00}{0.00} & \scorecell{0.00} & \scorecell{0.00} & \scorecell{0.00} & \scorecell{0.00} & \scorecell{0.00} & \scorecell{0.00} & \scorecell{0.00} & \scorecell{0.00} & \scorecell{0.00} & \scorecell{1.00} & \scorecell{1.29} & \scorecell{1.00} & \scorecell{1.00} \\
\bottomrule
\end{tabular}
}
\end{table*}

\end{document}